\documentclass[useAMS,usenatbib]{mn2e}
\usepackage[usenames,dvips]{color}
\usepackage{graphicx,natbib}
\usepackage{hyperref}
\usepackage{amsmath,amssymb}
\usepackage{multirow}
\usepackage{epsfig}

\newcommand{\eso}{ESO\,243$-$49 HLX$-$1}

\title[X-ray timing and spectral features of ESO 243-49 HLX-1]
{Investigation of X-ray timing and spectral properties of ESO 243-49 HLX-1 with long-term Swift Monitoring}       
\author[L. C.-C. Lin, C. P. Hu, K. L. Li, J. Takata, D. C. C. Yen, K. J. Kwak, Y. M. Kim and A. K. H. Kong]{Lupin Chun-Che Lin$^1$\thanks{E-mail:
lupin@unist.ac.kr}, Chin-Ping Hu$^{2\dagger}$, Kwan-Lok Li$^{1,3}$, Jumpei Takata$^4$,\and David Chien-Chang Yen$^5$,Kyujin Kwak$^1$, Young-Min Kim$^1$ and Albert K. H. Kong$^3$\\
$^1$Department of Physics, UNIST, Ulsan 44919, Korea\\
$^2$Department of Astronomy, Kyoto University, Kyoto 606-8502, Japan\\
$^3$Institute of Astronomy, National Tsing Hua University, Hsinchu 30013, Taiwan\\
$^4$School of Physics, Huazhong University of Science and Technology, Wuhan 430074, PRC\\
$^5$Department of Mathematics, Fu Jen Catholic University, New Taipei City 24205, Taiwan \\
$^{\dagger}$ JSPS International Research Fellow }

\begin{document}

\date{July 2019; ??? 2010}
\pagerange{\pageref{firstpage}--\pageref{lastpage}}
\pubyear{2019}
\maketitle
\label{firstpage}

\begin{abstract}
The long-term \emph{Swift} monitoring of \eso\ provides an opportunity to investigate the detailed timing and spectral behaviour of this hyper-luminous X-ray source. 
\emph{Swift} has detected 7 outbursts since 2009 mid-August.
Using different dynamical timing algorithms, we confirm an increasing trend for the time intervals between outbursts, which is manifest in the delays between the latest outbursts.
The X-ray spectra of HLX$-$1 in quiescence can be described with a single power-law model while the thermal component dominates the X-ray emission during outburst.
There is only marginal evidence for photon index (or spectral hardness) changes between quiescent states with about 1$\sigma$ deviation. 
%when measured count rates are obviously different.
With the updated temporal and spectral features, we re-examine different scenarios to explain the origin of the quasi-periodic modulation of HLX$-$1. 
A significantly increasing trend without obvious stochastic fluctuations on the timescale of the detected quasi-period may not fully support an orbital period origin 
as might be due to mass transfer episodes from a donor star at periastron of an extremely eccentric orbit.   
The outburst profile seems to be consistent with the effect of tidal-induced-precession of an accretion disc or an oscillating wind scenario in the inner disc.  
Based on these models, we speculate that the true orbital period is much shorter than the detected quasi-periodicity. 
      
\end{abstract}


\section{Introduction}
\label{sec:intro}
The hyper-luminous X-ray source with an isotropic X-ray luminosity of $\gtrsim 10^{42}$\,erg\,s$^{-1}$ \citep{Farrell2009,Godet2009} located in the edge-on galaxy ESO 243$-$49 is commonly recognized as an intermediate-mass black hole (IMBH of $10^2$-$10^5$\,$M_{\odot}$; \citealt{Makishima2000, MC2004}) candidate.
It is $\sim 8''$ north-east of the nucleus \citep{Farrell2009} of ESO 243$-$49, and its connection to the host galaxy was confirmed by the redshift measurement $z=0.0224$ from the H$\alpha$ emission of the optical counterpart \citep{Wiersema2010,SHP2013}. 
\eso\ (hereafter HLX$-$1), like other ultraluminous X-ray sources (ULXs) in nearby galaxies, was initially suggested as an IMBH based on the Eddington limit and the peak of the disc blackbody component detected, along with the spectral variability \citep{Farrell2009}. 
Nevertheless, some recent studies suspect the current identification and do not reject the possibility to treat a ULX as a stellar-mass black hole (BH; \citealt{Liu2013,Lasota2015}) or a neutron star \citep{Bachetti2014}.   

%TTTTTTTTTTTTTTTTTTTTTTTTTTTTTTTTTTTTTTTTTTTTTTTTTTTTTTTTT 
\begin{figure*}
\psfig{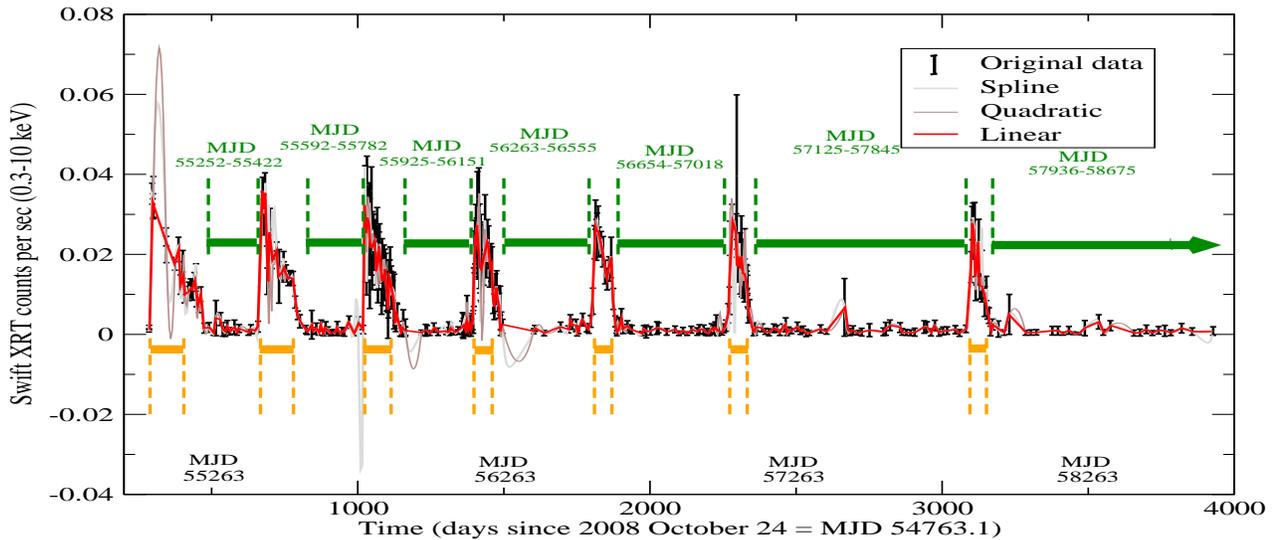}
\caption{\small{
Light curve of \eso\ obtained from \emph{Swift}/XRT in 0.3--10\,keV. 4 scattered data points before day 285 (MJD 55048) are removed in the related timing analysis. Solid lines of different colours are used to denote 5-d re-binned data obtained through different interpolation methods including linear, quadratic and spline interpolations. Only the data obtained via linear interpolation are considered in our timing analysis. The orange and green horizontal lines are used to label the time intervals of different outburst and quiescent epochs adopted for spectral analysis in Table~\ref{ev_spectra}}.   
}
\label{LC}
\end{figure*}
%%%%%%%%%%%%%%%%%%%%%%%%%%%%%%%%%%%%%%%%%%%%%%%%%%%%

The offset of $\sim 430$\,km\,s$^{-1}$ between the recession velocity for the nucleus of ESO 243$-$49 and that for HLX$-$1 suggests an indirect relation between them \citep{SHP2013}.
HLX$-$1 could belong to a dwarf galaxy \citep{Webb2010,MAZS2013I} or a globular cluster \citep{MC2004} near ESO 243$-$49.
If HLX$-$1 was at the center of a dwarf galaxy, it might have had a recent interaction with the host galaxy ESO 243$-$49 \citep{Farrell2012}.
%from \rmfamily H\,{\textsc i} radio observations \citep{Musaeva2015} and SPH (smoothed particle hydrodynamics) simulations \citep{MZM2012}.
The lack of \rmfamily H\,{\textsc i} emission indicates that the cluster environment depleted the gas during the merger process \citep{Musaeva2015}, and SPH (smoothed particle hydrodynamics) simulations confirmed the possibility of a merger event \citep{MZM2012,MAZS2013II}.
%However, more detailed parameters (i.e., mass of the supermassive BH centered in ESO 243$-$49, the age and metallicity of the cluster and the gas contents) are required to constrain a realistic simulation for the merger event to lead to the formation of an IMBH \citep{Webb2017}.

The X-ray telescope (XRT) of the \emph{Neil Gehrels Swift Observatory} (hereafter \emph{Swift}) provides a long-term monitor for the flux/spectral changes of HLX$-$1 to investigate its physical properties since the end of 2008 Oct. (i.e., $\sim$ MJD 54763.1).
A dramatic flux increase was detected in 2009 mid-Aug. \citep{Godet2009, Yan2015} and concluded as an outburst with a luminosity change by a factor of $\sim$50. 
The outbursts were found to be recurrent approximately with follow-up \emph{Swift} monitoring programs \citep{Lasota2011}.
However, an outburst of HLX$-$1 occurred in 2015 early January \citep{KSF2015} showed an extended interval of $\sim$ 460 d since the previous outburst, and the subsequent quiescent stage extended to more than 700 d until another high X-ray flux state was captured by \emph{Swift}/XRT in 2017 mid-April \citep{YY2017}.  
Each outburst in the X-ray light curve can be described with a ``fast-rise exponential-decay'' (FRED) profile, which is similar to outbursts detected from Galactic low-mass X-ray binary transients (LMXBTs; \citealt{CSL97,YY2015}).

The spectral state and the (1.5--10\,keV)/(0.3--1.5\,keV) hardness ratio (HR) also evolved following each outburst event.  
In the spectral analysis performed by \citet{Yan2015}, the spectral behavior can be mainly classified into two clusters, corresponding to the high/soft (HR $\leq$ 0.1) and low/hard (HR $\geq$ 0.6) states, which are also similar to those state transitions detected in Galactic LMXBTs \citep{Dunn2010,Munoz2014}.
More detailed spectral analyses were executed by \citet{TS2016II}, and the bulk motion Comptonization (BMC) model provided acceptable fits to all the spectral states.
The mass of HLX$-$1 was estimated as $M_{BH}\sim 7\times10^{4}$\,$M_{\odot}$ assuming a distance of 95\,Mpc by comparing the saturated photon index ($\Gamma$) versus the BMC normalization (proportional to the mass accretion rate $\dot{M}$) correlation with those of other Galactic BHs and extragalactic BHs.
Such a correlation also rules out identifying HLX$-$1 as a neutron star, as only the BH sources can show a changeable photon index ($\Gamma$) from 1.6 to 3 in the accretion process, while the photon index of a neutron star keeps a constant value close to $\Gamma \sim 2$ \citep{ST2011,Seifina2015}.
The inferred mass of HLX$-$1 fits with the BH mass range (i.e., $\sim 9\times10^{3} - 9\times10^{4}$\,$M_{\odot}$) given by the Eddington scaling through radio observations \citep{Webb2012}.  
The BH mass inferred from the correlation is also significantly heavier than a stellar-mass BH ($< 100$\,$M_{\odot}$) radiating at Eddington or super-Eddington rate (e.g., \citealt{Mukai2005}) or those detected with gravitational wave events \citep{GW2016,2018LIGO} due to BH mergers. 
%somewhat beyond the mass ($\sim 10^{4}$\,$M_{\odot}$) of a \MD{non}-spinning BH estimated from X-ray spectral model fitted with the variety of accretion disc \citep{Godet2012,Straub2014}.   

The most intriguing feature of HLX$-$1 is the highly variable interval between X-ray outbursts, which can be explained by modulated mass transfer due to tidal stripping of a companion star in an eccentric orbit \citep{Lasota2011}. 
The orbital delay in such a scenario can be reproduced by an SPH simulation \citep{Godet2014}.
An alternative explanation is a superorbital period due to a precession of the accretion disc \citep{Lin2015}.
We note that the outburst properties resulting from the disk instability model fail to account for the HLX$-$1 duty cycle timescales and the high X-ray luminosity obtained in the quiescent stages \citep{Soria2017}.
In comparison to the disk instability model, a strong inner oscillating wind scenario in a large accretion disc was also proposed to regulate the outburst of HLX$-$1 \citep{Soria2017} because the characteristic outer radius of a large disc of a few 10$^{13}$\,cm can be inferred from the optical observations \citep{Soria2012}.
After the latest report of outburst in 2015 Apr., we find one more outburst with a much longer delay using the \emph{Swift} archive.
We here provide complete timing and spectral analyses to investigate possible changes in the quasi-periodic signal and spectral behaviour.
%Based on the obtained results, we will further discuss their physical origin.
We describe the timing algorithms and spectral models in Sect. 2 and present the results in Sect. 3.
We discuss further constraints on possible physical mechanisms in Sect. 4.

\begin{table} 
\caption{Summary of \emph{Swift} observations used in analysis}\label{obs}
\begin{tabular}{ll} 
\hline \hline 
Obs. ID & Date of observations  
\\
\hline
\multicolumn{2}{l}{00031287 (005--170, 171--174, 176--179, 181--185, 187--220, 223--246, }
\\ 
248--252, 254--255)  & 2009 Aug. 5 -- 2012 Sep. 28 
\\ 
00032577 (001--187) & 2012 Oct. 2 -- 2019 Jul. 8 
\\ 
00049794 (001--004) & 2013 Mar. 14-- 2013 Mar. 22 
\\ 
00080013001 & 2012 Nov. 21 -- 2012 Nov. 21 
\\
00091907 (001--037) & 2014 Apr. 3 -- 2015 Mar. 15
\\
00092116 (001--029) & 2015 Apr. 15 -- 2015 Dec. 21
\\
00093143 (001--022) & 2017 Mar. 31 -- 2018 Mar. 19
\\
03104799 (002, 004--006) & 2018 Dec. 18-- 2019 Mar. 17
\\
\hline 
\hline 
\end{tabular}
\end{table}

\section{Data Reduction and Analysis}
\subsection{Observation and Data Reduction}
\label{sec:observation} 
All \emph{Swift} XRT data of HLX$-$1 were taken with the photon counting (PC) mode.
%We considered the \emph{Swift} data of \eso\ observed via the photon counting (PC) mode because a more precise source position can be determined without a serious contamination.
In order to perform timing analysis, we generated the light curve through the XRT products generator\footnote{http://www.swift.ac.uk/user\_objects} \citep{Evans2007, Evans2009}.
We extracted a circular source extraction region to match the point spread function, and only the source events in the effective energy range (i.e., 0.3--10\,keV) of \emph{Swift} with grades 0--12 are taken into account.
To correct for the background contribution, we used a large annulus region centered on our target.
The light curve is binned for each observation and is corrected for good time interval and bad pixel columns.

\emph{Swift} has monitored our target since the end of 2008 Oct., but only 4 scattered data points were obtained before 2009 early Aug. (i.e., MJD~55048). 
We therefore remove them in the subsequent timing analysis to avoid the impact that can be induced by large time gaps in the data.
We also reject data points contaminated by high background or spurious flares with anomalously high count rates, and we keep those data points with at least 3$\sigma$ detection significance.  
The 401 data points remaining were used to investigate the temporal behaviour of HLX-1, and contain at least 7 significant outbursts (as shown in Fig.~\ref{LC}).
In order to employ specific dynamical timing algorithms, which require evenly sampled data, we also re-binned the data every 5 d and then applied different interpolation methods (e.g., linear, spline and quadratic) to fill the gaps.  
As shown in Fig.~\ref{LC}, we can see that spline or quadratic interpolations lead to artificial peaks in the light curve and will cause fake signals, so we only used the linear interpolation to generate the evenly sampled data sets for timing analysis.  

We list all the \emph{Swift} data used in timing and spectral analysis in Table~\ref{obs}. 
%We considered the data with our target inside the field-of-view, including Obs. ID 00032577(101--187; from 2015 Dec. 22 to 2019 Jul. 8), 00093143(001--022; from 2017 Mar. 31 to 2018 Mar. 19), 03104799(002, 004--006; from 2018 Dec. 7 to 2019 Mar. 17), 07016068001 (2019 Apr. 25), 07016069001 (2019 Apr 25) and 07016431001 (2019 Apr 26).
In the process to derive HRs and to generate related spectral products of HLX$-$1, we used HEASOFT v6.22.1, with the task ``xrtgrblc v1.9'' and the latest calibration files (CALDB ver. is 20180710\footnote{https://heasarc.gsfc.nasa.gov/docs/heasarc/caldb/swift/}).
The exposure maps and ancillary response files were also generated using the task ``xrtgrblc''.
The response matrix files were adopted accordingly in our spectral fit. 
%and we also independently checked our \RE{results} in comparison to those obtained from the online XRT data product generator.
We combined those \emph{Swift} data with a similar hardness ratio (HR) to generate a spectrum and fitted it with XSPEC v12.9.1p\footnote{http://www.heasarc.gsfc.nasa.gov/xanadu/xspec/manual/manual.html}.

%Each grouped spectrum was rebinned to a minimum counts of 15 counts per channel to ensure the $\chi^2$ statistic in our fitting.   

\subsection{Timing algorithms}
\label{sec:timing analysis}
\begin{figure}
\psfig{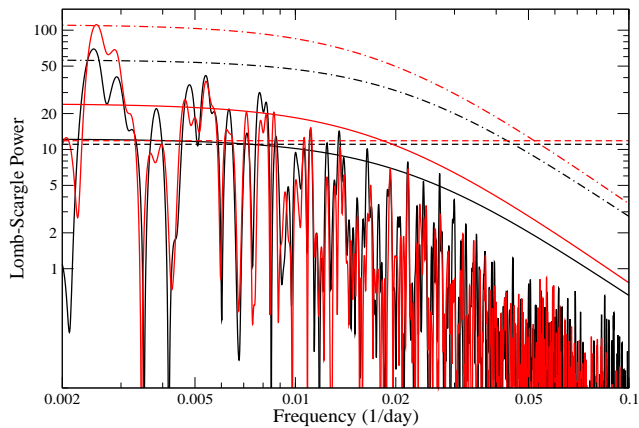}
\caption{\small{LSP of \eso. The black and red curves denote the bias-corrected LSPs of the original data set and the re-sampled data with 5-d binning. The red noise models of each LSP obtained from the original and re-sampled data sets are indicated with the black and red solid lines, respectively. The 99\% white and red noise significance levels of the original (in black) and re-sampled light curves (in red) are represented with dashed horizontal lines and dash–dotted curves.}   
}
\label{LSP}
\end{figure}

In the preliminary investigation of the timing signal, we used the Lomb-Scargle method \citep{Lomb76,Scargle82}, which was improved from the discrete Fourier transform to deal with unevenly sampled data sets.
We generated the periodogram (i.e., LSP) via REDFIT v3.8e\footnote{https://www.marum.de/Michael-Schulz/Michael-Schulz-Software.html} \citep{SM2002} and considered the effects of both white and red noise.
In Fig.~\ref{LSP}, we show the LSP obtained from the original data sets and a 5-d average data, which were corrected for the bias on the overestimation of the high-frequency power due to the uneven temporal sampling in time domain \citep{SS1997,SM2002}.
In order to cover the frequency range to 0.1 d$^{-1}$ with enough resolution, we set the oversampling factor for the Lomb-Scargle method and the factor of the maximum frequency relative to the Nyquist frequency in our analysis as 64 and 2, respectively.
%We also translated the amplitude of the obtained spectrum into Lomb-Scargle power with a scaling factor in relation to the average sampling rate \RE{and the count rate variance}. 
The white noise false alarm probability was derived according to the power of the highest peak in a periodogram ($P$) and the number of independent trial frequencies ($N_i$) in the computation (i.e., $1-(1-e^{-P})^{N_i}$), where the number of independent trials is obtained from the Horne number \citep{HB86}.
The red noise model can be estimated through a fit to the first-order autoregressive (AR1) process to the light curve \citep{SM2002}, and the false alarm levels were determined by scaling the red-noise model with an appropriate percentile of the $\chi^2$-probability distribution (see solid lines in Fig.~\ref{LSP}).

%TTTTTTTTTTTTTTTTTTTTTTTTTTTTTTTTTTTTTTTTTTTTTTTTTTTTTTTTT 
%%%%%%%%%%%%%%%%%%%%%%%%%%%%%%%%%%%%%%%%%%%%%%%%%%%%

%TTTTTTTTTTTTTTTTTTTT  
\begin{figure*}
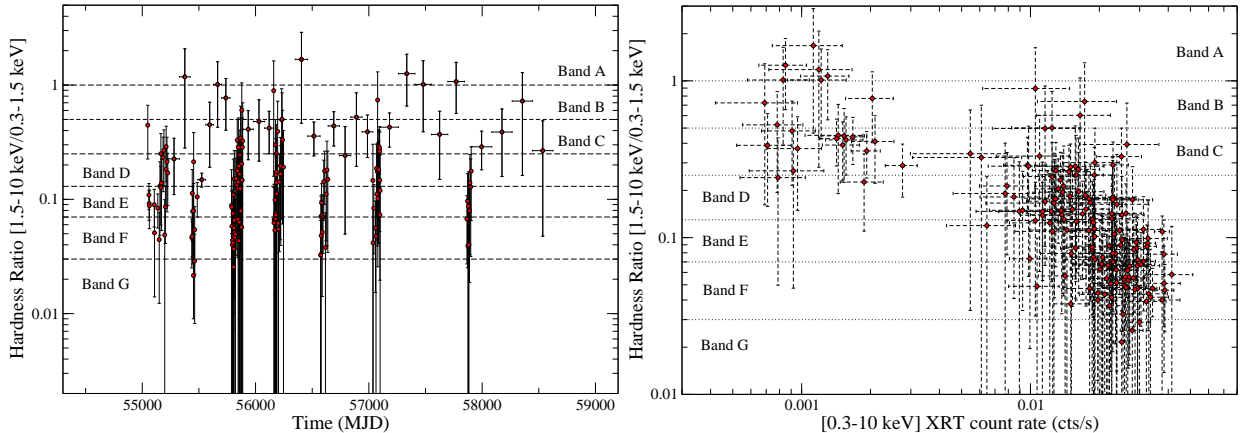
 \centering
\psfig{file=newHR-time.eps,width=3.2in}
\psfig{file=newHR-cts.eps,width=3.2in}
\caption{(Left panel): Evolution of hardness ratio for \eso from 2009 Aug. to 2019 Jun. determined by the \emph{Swift} data. To have enough source photons for a statistical investigation, we sum the data in quiescence into longer time intervals. In the obvious outburst intervals, we determine HR from individual observations.  (Right panel): Hardness-intensity diagram for \eso  from 2015 Apr. to 2019 Apr determined by the \emph{Swift} data.} 
\label{HR}
\end{figure*}
%TTTTTTTTTTTTTTTTTTTT  
%\citet{Clarkson2003a,Clarkson2003b} 
We further trace the variability of the frequency of the quasi-periodic signal using the dynamic power spectrum (DPS), i.e., dividing the entire time interval into several sliding windows and calculate the LSP of each of them.
The idea of such a dynamical power spectrum (DPS) comes from \citet{Clarkson2003a,Clarkson2003b}, and the variation of the power for the quasi-periodic signal can be investigated in different time intervals.
We set the length of the sliding window as 1000\,d to cover at least two complete cycles of the most powerful major signal that can be roughly estimated by the average separation of early outbursts (i.e., $\sim$380\,d before early 2014; \citealt{Lin2015}).
However, the outburst recurrence intervals seem to extend, and the quasi-period of the main signal can be significantly longer than the earlier detection if we include data sets after the 6th detected outburst (i.e, $\sim$MJD 57120; 2015 April).
Furthermore, we apply decomposition methods to trace how the major signal evolved.
The WWZ (weighted wavelet Z-transform; \citet{Foster96}) and HHT (Hilbert-Huang transform; \citealt{Huang98}) are the decomposition methods that are widely adopted to investigate long-term variation/periodicities (e.g., WWZ of \citet{TLW2016} and HHT of \citet{Hu2011}).
WWZ is implemented by casting the wavelet transform as a projection, and it consists of the effective number and the weighted variations of uneven data and model functions so it is optimized for non-stationary unevenly sampled time series analysis \citep{Han2012, Lin2015}.

Compared with WWZ, HHT can only be employed on evenly sampled data, and we applied it to the 5-d average light curve. 
HHT decomposes the original time series into several intrinsic mode functions (IMFs) to represent the oscillatory mode as a counterpart to the harmonic function, where the component of the largest energy/power can correspond to the main signal embedded in the light curve.
In our analysis using HHT, we considered an ensemble empirical mode decomposition (EEMD; \citealt{Wu2009}) to avoid the mode mixing problem and decomposed the original light curve into 10 IMFs.    
After decomposing the original light curve into IMFs, the normalized Hilbert transform \citep{HL2003} can be applied to the IMFs to obtain the instantaneous frequencies and amplitudes.
The extraction of instantaneous frequencies and amplitudes of different IMFs can display how the modulation period and amplitude vary with time on the the resultant Hilbert spectrum, and we determined the Hilbert energy/power as the square of the instantaneous amplitude. 
The Hilbert spectrum can show the frequency and energy of each IMF as functions of time, and such an algorithm can be used to investigate the change of each cycle length (i.e., inter-wave modulation) and the instantaneous frequency in one cycle (i.e., intra-wave modulation).
%Based on the features of these dynamical algorithms, the wavelet transform \citep{CBMK2004} and the HHT \citep{CCN2007,Sakai2017} are also applied to detect the gravitational wave (GW).
%The most common GW event trigger generator, \emph{Omicron}, is developed based on Q-transform \citep{Lynch2017}, which is also closely related to the Gabor-Morlet wavelet transform \citep{LW93}.
%\emph{$\eta$Gen} based on HHT is a new GW event trigger generator developed by KGWG (Korean Gravitational Wave Group), and it demonstrates a comparable performance to \emph{Omicron} \citep{Son2018}. 
%The aforementioned applications clearly show the capability of these dynamical algorithms to check out the signal. 
%The quasi-periodic signal we want to detect for HLX$-$1 is relatively stable and does not cross a wide frequency range such as the compact binary coalescence.
%In comparison to \RE{the analyses for} GW, it is not necessary to vary the Q-value to reveal the main signal in our study; however, WWZ that consists of more sinusoidal components can provide a more detailed signal pattern than the constant-Q transform.
The constant-Q transform \citep{Brown91} can be another useful algorithm for signals with exponential growth in frequency; however, WWZ consists of more sinusoidal components and provides a more detailed signal pattern than the constant-Q transform. 
The reason is that the WWZ resolves resolves the frequency range linearly, whereas the constant-Q transform resolves it in a geometric scale of factor Q
Therefore, we only performed the dynamical temporal analysis with WWZ and HHT for this case with limited data points. 
  
\subsection{Hardness-intensity and spectral analysis}
\label{sec:spectral analysis}
%The HR and spectra of the first 6 outbursts in the X-ray band have been well studied in the previous literatures \citep{Yan2015,TS2016II}.
To investigate the evolution of the spectral state after the 6th outburst (i.e., $\sim$MJD 57120), we included recent \emph{Swift} data to check both HR and spectral behavior.
We use HR as the ratio of hard (1.5--10\,keV)/soft (0.3--1.5\,keV) counts and classify it into 6 bands from low to high (i.e., `A+B': HR $>0.5$;  `C': $0.25 <$ HR $< 0.5$; `D': $0.13<$ HR $<0.25$; `E': $0.07<$ HR $<0.13$; `F': $0.03<$ HR $<0.07$; `G': HR $<0.03$) to compare with the spectral states seen before the 6th outburst \citep{TS2016II}.
We note that the HR determined from a single observation in the quiescent stage has a very large uncertainty due to limited source photons.
Thus, we combined several observations (i.e., 8--16) in quiescence to accumulate enough source photons to determine a HR.
In the outburst stage, we determine the HR from each single observation because the count rate is at least one order of magnitude larger than that in quiescence.   
The evolution of HR and the HID (hardness-intensity diagram) is displayed in Fig.~\ref{HR}. 
We can clearly see that the 7th outburst occurred around MJD 57862 (i.e.,  2017 mid. Apr.), and \eso\ obviously underwent a spectral transition from the low/hard to high/soft state during the outburst. 
The possible flare seen at $\sim$MJD 57429 stayed in the hard spectral state and obviously had a different HR compared with the outburst stage.

%TTTTTTTTTTTTTTTTTTTT  
\begin{figure*} \centering
\psfig{file=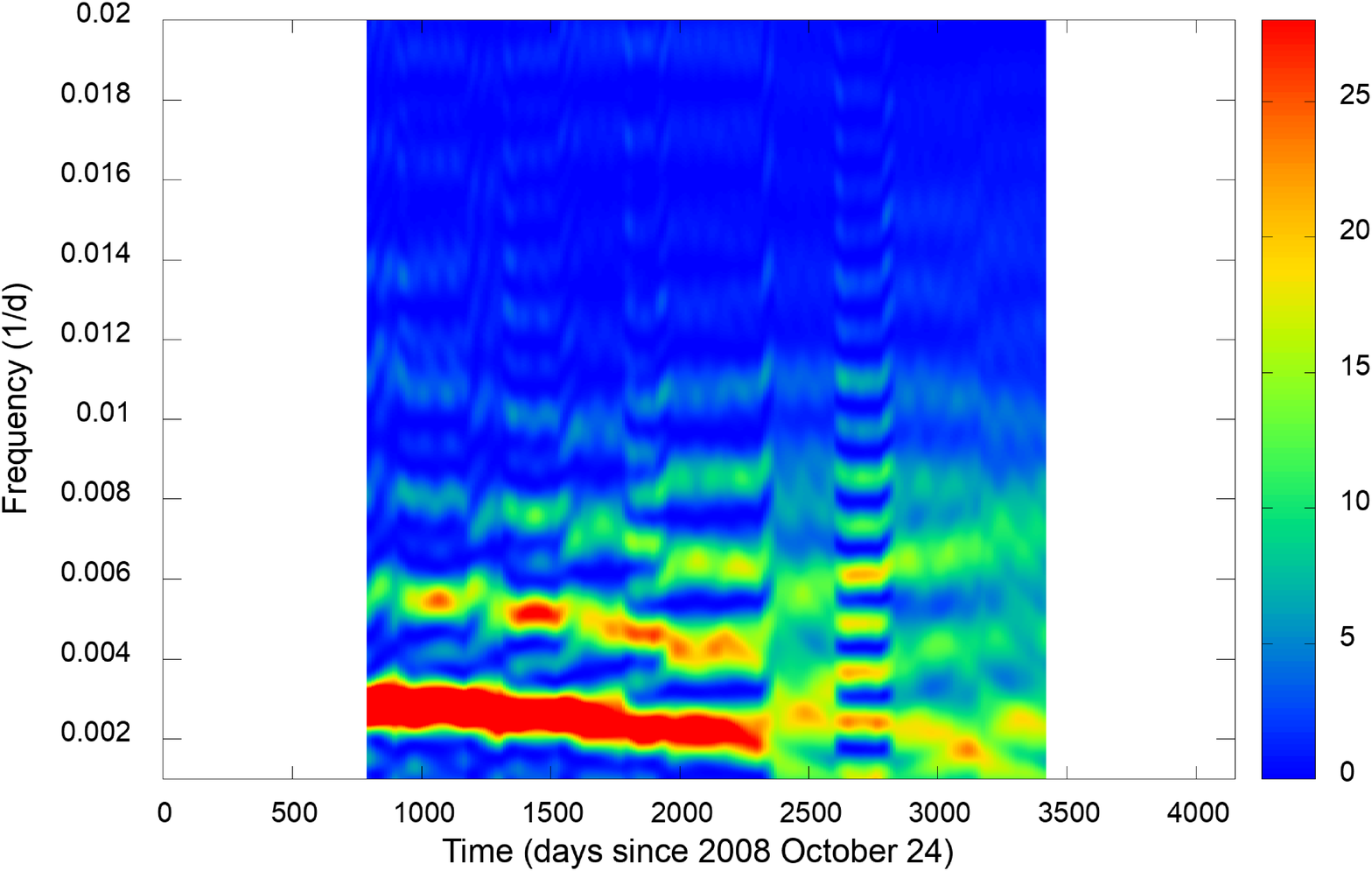,width=3.6in}
\psfig{file=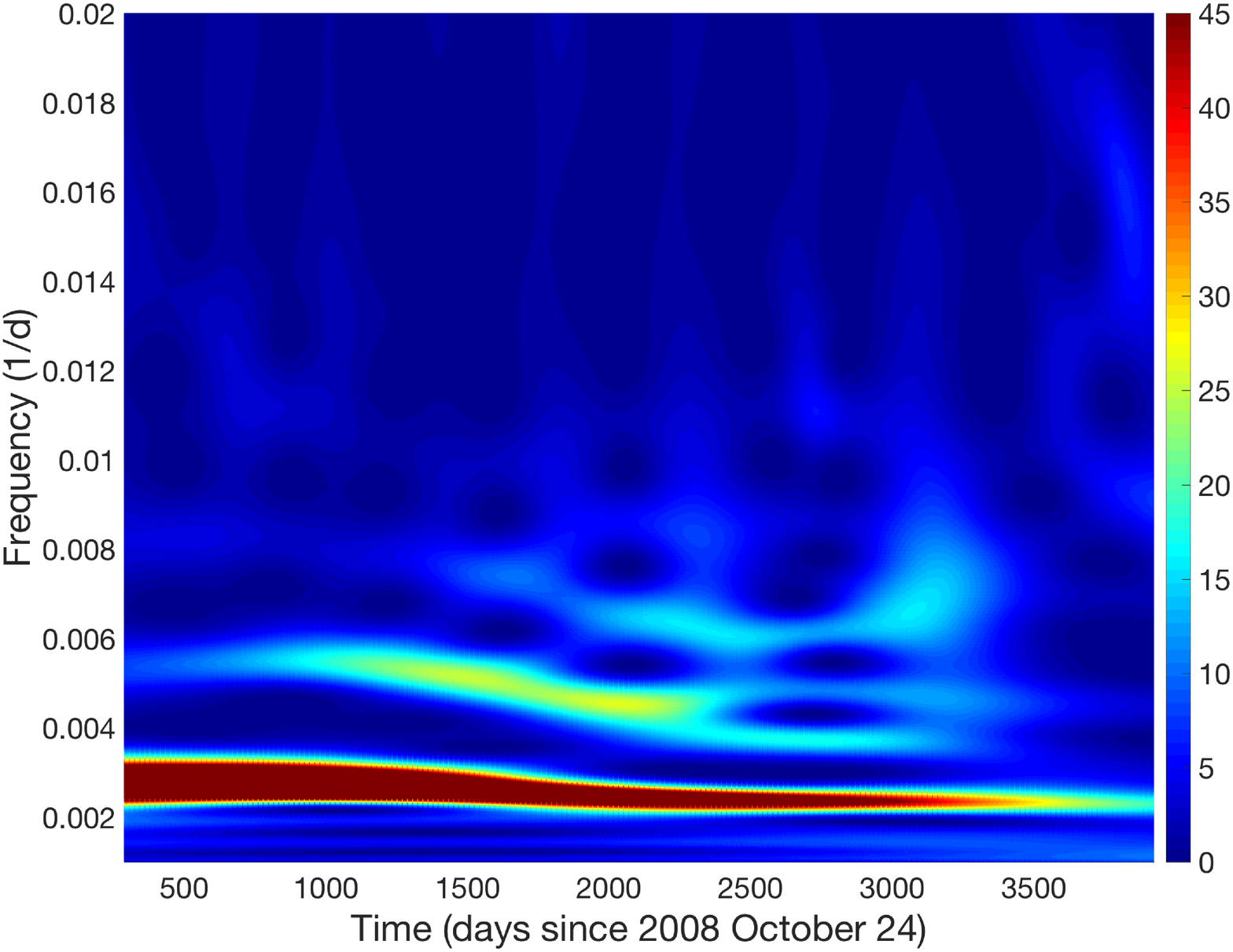,width=3.3in}
\caption{(Left panel): Dynamical power spectrum (DPS) of HLX$-$1 from rebinned data with a moving window of 1000\,d and moving step size of 10\,d.  (Right panel): WWZ periodogram of HLX$-$1 from rebinned data. In both cases contour levels of power obtained are shown in the colour scales.}
\label{periodogram}
\end{figure*}
%TTTTTTTTTTTTTTTTTTTT  

Due to a low X-ray count rate and limited exposures of each observation, we do not have enough source photons to perform spectral analysis with a single observation. 
We therefore accumulated source events of different observations labelled with the same HR band to generate a spectrum.
The source photons accumulated in the high HR band (i.e., A+B) for the spectral analysis are few, so we employed the Cash statistic \citep{Cash79} for spectral fitting.
In the other HR bands, we regrouped channels to have at least 15 source photons in each channel to allow use of $\chi^2$ statistic. 
We applied interstellar absorption using Wisconsin cross-sections \citep{MM83} and fixed the neutral column as $5\times 10^{20}$\,cm$^{-2}$ \citep{Yan2015}.
We performed spectral fits using a single power-law (PL) model first.
We also considered multi-component models such as PL+BB (blackbody)/discBB (multi-colour disc; an accretion disk consisting of multiple blackbody components) or BMC (Bulk Motion Comptonization; \citealt{TMK97}).
The log$A$ in the BMC model is introduced to evaluate the contribution from the illumination of in-falling matter, and it is fixed at -0.3 to denote the 67\% BB emission of the Comptonization spectrum that is visible to us and the other 33\% BB emission is convolved with the Comptonization Green function when its uncertainty cannot be reasonably constrained in the best fit.

%TTTTTTTTTTTTTTTTTTTT  
\begin{table*}
\caption{Best-fit spectral parameters for \eso\ in different \emph{Swift} HR bands.}\label{spectrum}
\begin{tabular}{clcccccc} 
\hline
\hline
 &  & A+B Band & C Band & D Band & E Band & F Band & G Band
\\
\hline
\multicolumn{2}{c}{Hardness ratio} & $> 0.5$ & 0.25--0.5 & 0.13--0.25 & 0.07--0.13 & 0.03 -- 0.07 & $< 0.03$
\\
\hline
\multicolumn{2}{c}{Model/Parameters} &  &  &  &  &  &
\\
\hline
 & $\Gamma_{\rm{PL}}$ & 1.6$^{+0.3}_{-0.2}$ & 2.7$\pm 0.3$ & 3.15$\pm 0.09$ & $3.26\pm 0.08$  & $^{g)}$3.5 & 3.1$^{+0.2}_{-0.1}$
\\
PL & $^{a)}N_{\rm{PL}}$  & 1.2$\pm 0.3$ & 2.7$\pm 0.4$ & 7.0$\pm 0.5$ & 14.7$\pm 0.9$ & $^{g)}$13.6 & 17.3$^{+1.8}_{-1.7}$
\\
  & $^{b)}F_{\rm{PL}}$  & 0.9 & 1.3 & 3.8 & 8.4 & 8.6 & 9.3
\\
 & $\chi^2$/d.o.f. &  $^{f)}$78.0/68  & 33.0/29 & 71.7/62 &120.3/73 & 190.5/74 & 50.6/42
\\
\hline
  & $\Gamma_{\rm{PL}}$ & 1.2$^{+0.5}_{-0.4}$ & 1.9$\pm 0.4$ & 2.6$^{+0.4}_{-0.5}$ & 2.9$^{+0.3}_{-0.4}$ & 3.5$^{+0.9}_{-0.6}$ & --- 
\\
PL & $^{a)}N_{\rm{PL}}$  & 0.8$^{+0.8}_{-0.4}$ & 1.8$^{+0.5}_{-0.6}$ & 4.1$^{+1.2}_{-1.4}$ & 7.2$^{+1.8}_{-2.0}$ & 4.3$^{+1.7}_{-2.2}$ & ---
\\
+  & $kT_{\rm{BB}}$ (eV) & 124$^{+51}_{-47}$ & 113$^{+18}_{-19}$ & 134$^{+16}_{-11}$ & 147$^{+12}_{-9}$ & 154$^{+13}_{-12}$ & 147$\pm 7$
\\
BB & $^{c)}N_{\rm{BB}}$ & 0.3$^{+0.4}_{-0.2}$  &  0.9$^{+0.4}_{-0.3}$  & 2.5$^{+0.9}_{-1.1}$ & 6.2$^{+1.7}_{-1.5}$  & 6.5$^{+1.5}_{-1.2}$  & 9.6$^{+1.0}_{-1.0}$
\\
  & $^{b)}F$  & 1.1 & 1.6 & 3.6 & 7.8 & 7.4 & 6.5
\\
 & $\chi^2$/d.o.f. & $^{f)}$73.6/66 & 20.5/27 & 48.6/60 & 48.2/71 & 84.1/72  & 51.4/42
\\
\hline
  & $\Gamma_{\rm{PL}}$ & 1.6$^{+0.3}_{-0.2}$ & 1.9$^{+0.4}_{-0.5}$ & 2.1$^{+0.6}_{-0.7}$ & 2.4$^{+0.6}_{-1.2}$  & --- & --- 
\\
PL & $^{a)}N_{\rm{PL}}$  & 0.8$^{+0.4}_{-0.3}$ & 1.7$\pm 0.6$ & 2.4$^{+1.8}_{-1.3}$ & 4.0$^{+2.7}_{-2.8}$ & --- & ---
\\
+  & $kT_{\rm{in}}$ (eV) & 151$^{+108}_{-76}$ & 150$\pm 33$ & 187$\pm 17$ &  201$^{+13}_{-12}$ & 207$^{+7}_{-8}$ & 204$\pm 12$
\\
discBB  & $^{d)}R_{\rm{in}}$ ($10^4$\,km) & 1.7$^{+2.8}_{-1.3}$  &  3.3$^{+2.5}_{-1.3}$  & 3.8$\pm 0.8$ & 5.1$^{+4.3}_{-1.1}$  & 5.2$^{+0.6}_{-0.5}$  & 5.4$^{+0.9}_{-0.8}$
\\
  & $^{b)}F$  & 1.1 & 1.6 & 3.7 & 7.9  & 7.0 & 7.1
\\
 & $\chi^2$/d.o.f. & $^{f)}$73.9/66 & 21.4/27 & 47.1/60 & 49.2/71 & 85.7/74  & 41.2/42
\\
\hline
  & $\Gamma_{\rm{BMC}}$ & 1.2$^{+0.4}_{-0.2}$ & 1.9$\pm 0.4$ & 2.7$^{+0.7}_{-0.6}$  & 3.7$^{+0.9}_{-0.8}$  & 4.4$^{+0.8}_{-0.5}$ & 2.2$^{+3.4}_{-1.1}$ 
\\
  & $kT_{\rm{s}}$ (eV) & 118$^{+51}_{-50}$ & 110$^{+18}_{-19}$ & 121$^{+13}_{-20}$ & 123$^{+13}_{-19}$ & 132$\pm 7$ & 121$^{+15}_{-42}$
\\
BMC & log$A$ & 0.05$^{+0.63}_{-0.54}$ &  -0.38$^{+0.26}_{-0.19}$  & -0.38$\pm 0.64$  & -0.06$^{+0.06}_{-0.52}$  & $^{e)}-0.3$ & -0.57$^{+0.60}_{-0.30}$
\\
  & $^{c)}N_{\rm{BMC}}$  & 0.6$^{+5.6}_{-0.2}$ &  1.4$^{+0.4}_{-0.2}$ & 4.3$\pm 0.4$ & 9.9$^{+0.7}_{-0.5}$  & 9.8$^{+0.8}_{-0.9}$ & 10.4$^{+50.8}_{-1.6}$
\\
  & $^{b)}F_{\rm{BMC}}$ & 1.1 & 1.6 & 3.5 & 7.5 & 6.9 & 9.4
\\
 & $\chi^2$/d.o.f. & $^{f)}$73.5/66 & 20.3/27 & 50.2/60 & 46.2/71 & 87.7/73  & 38.7/40
\\
\hline
\hline
\end{tabular}
%}
\begin{small}
\begin{flushleft}  
{\footnotesize
All fits are obtained with a fixed absorption ($N_H$) of 5.0$\times 10^{20}$ cm$^{-2}$, and all uncertainties represent a 90\% confidence interval for one parameter of interest.\\
$^{a)}$ Normalisation factor in units of $10^{-5}$\,photons~keV$^{-1}$\,cm$^{-2}$\,s$^{-1}$ at 1\,keV.\\
$^{b)}$ Unabsorbed flux (0.3--10\,keV) in units of $10^{-13}$\,ergs\,cm$^{-2}$\,s$^{-1}$.\\
$^{c)}$ Normalisation factor in units of $L_{33}/d^2_{10}$\,ergs\,s$^{-1}$\,kpc$^{-2}$, where $L_{33}$ is in units of $10^{33}$\,ergs\,s$^{-1}$ and $d_{10}$ is in units of 10\,kpc.\\
$^{d)}$ Inner disc radius assuming d=95\,Mpc and face-on disc.\\
$^{e)}$ Uncertainties cannot be reasonably constrained, so logA fixed at -0.3 (see text for details).\\
$^{f)}$ From Cash statistics.\\
$^{g)}$ Fit too poor to yield uncertainties.
%The log$A$ is fixed as -0.3 to denote the 67\% BB emission of the Comptonization spectrum is visible for the Earth observer and other 33\% BB emission is convolved with the Comptonization Green function when its uncertainty cannot be reasonably constrained in the best fit.\\
}
\end{flushleft}
\end{small}
\end{table*}
%TTTTTTTTTTTTTTTTTTTT
    
\section{Results}
\label{sec:result}   

\subsection{Signals determined by timing analysis}
\label{sec:timing result}

In Fig.~\ref{LSP}, we can see a double-peaked structure in the major signal and sub-signals.
This may indicate a change of the periodicity because the strong power of a trial frequency obtained at different time intervals can simultaneously project on a static LSP.
If we consider the strongest power of the LSP peak determines the period of the major and sub- signals, they correspond to $396\pm 2$\,d, $184.5\pm 0.8$\,d and $127.7\pm 0.4$\,d.
The uncertainties of the aforementioned quasi-periods are determined by the error propagation from two approaches to avoid an underestimation.  
We consider eq. (3) of \citet{Levine2011} as one method to determine the uncertainty of the frequency for a trial signal. 
We also estimated the uncertainty through a Monte Carlo simulation by randomly generating $10^4$ light curves assuming a Gaussian distribution associated with the original data.
We then repeated the timing analysis to derive a standard deviation for the peak in the frequency domain.
We note that the Fourier width of these signals correspond to 21.5\,d, 4.7\,d and 2.2\,d, which are much longer than the determined uncertainties.
Compared to a periodicity of 378$\pm$6\,d obtained in \citet{Lin2015}, we now identify the strongest peak to be at $396\pm2$\,d from our more extended database.
   
Fig.~\ref{periodogram} presents the evolution of signals that can be resolved from the data. 
The DPS has a lower resolution than the WWZ, but both of them demonstrate similar components. 
On the DPS of Fig~\ref{periodogram}, the major signal significantly decreased its power at t$\sim$2300\,d, which roughly corresponds to the occurrence of the 6th outburst.
We chose a small kernel parameter (i.e., c $\sim$0.005) so that the exponential term can relatively smoothly decay in each computing cycle for wavelet analysis to reach better spectral resolution than DPS.  
We can clearly see 3 signals in the WWZ map.
All of these signals corresponding to frequencies between 0.002--0.003, 0.004--0.006, and 0.006--0.008\,d$^{-1}$ can also be seen in the static LSP of Fig.~\ref{LSP} and the DPS in the left panel of Fig.~\ref{periodogram}, but the signal pattern is more blurred in the DPS.
The main signal with the strongest power gradually decreased its frequency, and the periodicity of the signal potentially represents the recurrence time of an outburst because we can also see the extension of the outburst interval in the light curve.

\begin{figure*} \centering
\psfig{file=IMFs.eps,width=3.1in}
\psfig{file=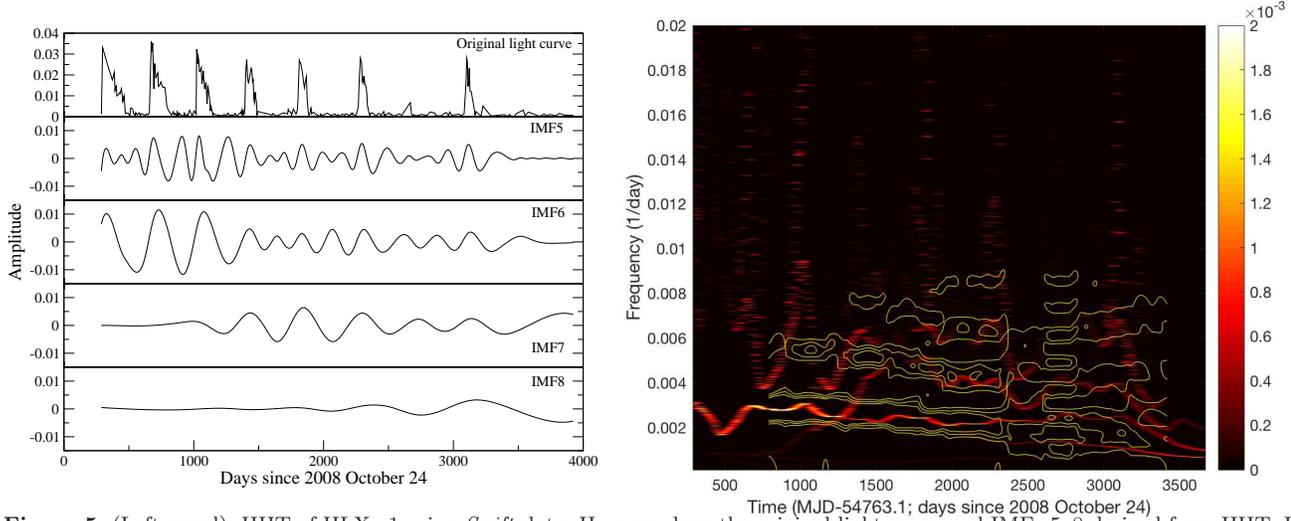,width=3.8in}
\caption{(Left panel): HHT of HLX$-$1 using \emph{Swift} data. Here we show the original light curve and IMFs 5--8 derived from HHT. IMFs 5--8 dominate the most significant Hilbert energy/power shown on the Hilbert spectrum. (Right panel): DPS superimposed on Hilbert spectrum for HLX$-$1. We show contours of the DPS obtained from the rebinned light curve and overlay it on the Hilbert spectrum.} 
\label{HHT}
\end{figure*}

In the WWZ periodogram, we find that sub-signals between 0.004--0.008\,d$^{-1}$ show a similar pattern to that displayed in the DPS; however, we do not find any bifurcation feature to connect to the main signal on the WWZ map.
In Fig.~\ref{LSP}, we can determine the timescale of the quasi-periodicities for those sub-signals, and it seems that they are harmonics of the main signal.
If we further sinusoidally fit the light curve to the sub-signal with a fixed timescale, we find the quasi-periodicities of $\sim$90--200\,d for the sub-signal represent the duration/duty cycle timescale of an outburst \citep{Lin2015} at different time intervals. 
We also note that the outburst duration decreased when it recurred \citep{Godet2014}, which also leads to the periodicity change of the most prominent sub-signal in the periodogram.
Except for a major signal corresponding to the frequency of $\sim 0.0253$\,d$^{-1}$ and sub-signals corresponding to frequencies $\gtrsim 0.004$\,d$^{-1}$, we note that a weak signal with a frequency lower than 0.002\,d$^{-1}$ in the WWZ periodogram enhanced its significance recently.
Such a signal cannot be clearly seen in the DPS because of the widow size, but a marginal indication for the bifurcation of the major signal to denote the change of the outburst interval is obvious after the 6th outburst.   
In the Hilbert spectrum shown in the right panel of Fig.~\ref{HHT}, we can also see that this low-frequency signal has become more prominent after t$\sim$2300\,d.
The reinforcement of this signal can be related to a significant change in the outburst interval because the 7th outburst was detected at t$\sim$3100\,d i.e., 2018 mid-April after being in quiescence for almost two years \citep{YY2017}.
It has an obvious extension compared with the time interval between the previous outbursts from $\sim 370-460$\,days \citep{KSF2015}.  

%TTTTTTTTTTTTTTTTTTTT  
%TTTTTTTTTTTTTTTTTTTT  

\begin{table*}
\caption{Best-fit spectral parameters for \eso\ determined from \emph{Swift} data at different stages.}\label{ev_spectra}
\begin{tabular}{clccccccc} 
\hline
\hline
\multicolumn{2}{c}{} & 1 & 2 & 3 & 4 & 5 & 6 & 7 
\\
\hline
\multicolumn{2}{c}{*Quiescent stage in MJD} & 55252--55422 & 55592--55782 & 55925--56151 & 56263--56555 & 56654--57018 & 57125--57845 & 57936--58675
\\
\hline
\multicolumn{2}{c}{Hardness ratio} & 0.38$\pm 0.22$ & 0.77$\pm 0.25$ & 0.30$\pm 0.11$ & 0.53$\pm 0.18$ & 0.54$\pm 0.15$ & 0.65$\pm 0.14$ & 0.38$\pm 0.12$ 
\\
\hline
 & $\Gamma_{\rm{PL}}$ & 2.5$\pm 0.4$ & 1.5$\pm 0.3$ & 2.5$^{+0.4}_{-0.3}$ & 2.2$\pm 0.3$ & 2.1$^{+0.3}_{-0.2}$ & 1.6$\pm 0.2$ & 2.0$\pm 0.3$
\\
PL & $^{a)}N_{\rm{PL}}$  & 2.0$^{+0.5}_{-0.4}$ & 1.1$^{+0.3}_{-0.2}$ & 1.3$^{+0.3}_{-0.2}$  & 1.4$^{+0.3}_{-0.2}$& 1.2$\pm 0.2$ & 1.1$\pm 0.2$ & 1.7$\pm 0.3$
\\
  & $^{b)}F_{\rm{PL}}$  & 1.0 & 1.0 & 0.6 & 0.7 & 0.6 & 0.8 & 1.0
\\
 & C-stat/d.o.f. & 7.6/12 & 19.0/17 & 19.7/24 & 19.7/16 & 47.9/38 & 49.5/42 & 9.3/19
\\
\hline
\multicolumn{2}{c}{*Outburst stage in MJD} & 55053--55168 & 55430--55543 & 55787--55877.5 & 56161--56223 & 56573--56633 & 57036--57096 & 57858--57915
\\
\hline
\multicolumn{2}{c}{Hardness ratio} & 0.086$\pm 0.011$ &  0.077$\pm 0.009$ & 0.066$\pm 0.008$  & 0.099$\pm 0.018$ & 0.079$\pm 0.015$ & 0.144$\pm 0.023$ & 0.067$\pm 0.016$
\\
\hline
 & $\Gamma_{\rm{PL}}$ & 2.8$\pm 0.3$ & 3.0$\pm 0.3$ & 3.2$\pm 0.2$ & 3.1$\pm 0.1$ & --- & 2.2$\pm 0.4$ & ---
\\
PL & $^{a)}N_{\rm{PL}}$  & 6.5$^{+1.4}_{-1.5}$ & 4.8$^{+0.9}_{-1.0}$ & 5.4$^{+0.8}_{-0.9}$ & 13.6$\pm 0.9$ & --- & 4.2$^{+1.6}_{-1.5}$ & ---
\\
+ & $kT_{\rm{BB}}$ (eV) & 153$^{+9}_{-8}$ & 144$^{+7}_{-5}$ & 147$^{+7}_{-6}$ & --- & 156$\pm 5$ & 128$^{+7}_{-8}$ & 157$^{+6}_{-7}$
\\
BB & $^{c)}N_{\rm{BB}}$ & 6.1$^{+1.1}_{-1.2}$ & 6.0$^{+1.0}_{-0.9}$ & 5.3$^{+0.9}_{-0.8}$ & --- & 8.9$\pm 0.5$ & 5.5$^{+0.9}_{-1.1}$ & 6.9$\pm 0.5$ 
\\
 & $^{b)}F$ & 7.4 & 6.5 & 6.7 & 7.3 & 6.2 & 5.6 & 4.8
\\
 & $\chi^2$/d.o.f. & 38.8/41 & 66.3/55 & 51.8/55 & 29.3/19 & 23.3/22 & 25.0/17 & 17.9/16
\\ 
\hline
 & $\Gamma_{\rm{PL}}$ & 2.1$^{+0.6}_{-1.2}$ & 1.9$^{+0.7}_{-1.1}$ & 2.8$^{+0.4}_{-0.5}$ & 2.6$^{+0.7}_{-2.5}$ & --- & 2.1$^{+0.4}_{-0.5}$ & ---
\\
PL & $^{a)}N_{\rm{PL}}$  & 3.0$^{+2.1}_{-2.2}$ & 1.5$^{+1.5}_{-1.0}$ & 3.2$^{+1.2}_{-1.4}$ & 8.4$^{+3.2}_{-4.3}$ & --- & 3.4$^{+1.8}_{-1.7}$ & ---
\\
+ & $kT_{\rm{in}}$ (eV) & 210$^{+11}_{-8}$ & 196$\pm 7$ & 195$^{+8}_{-7}$ & 172$^{+86}_{-25}$ & 215$^{+8}_{-9}$ & 175$\pm 14$ & 215$\pm 10$
\\
discBB & $^{d)}R_{\rm{in}}$ ($10^4$\,km) & 4.6$^{+0.5}_{-0.6}$ & 5.4$\pm 0.4$ & 5.0$\pm 0.6$ & 5.3$^{+3.0}_{-3.7}$ & 4.6$\pm 0.5$& 5.7$^{+1.0}_{-0.9}$ & 4.1$\pm 0.5$ 
\\
 & $^{b)}F$ & 7.6 & 6.7 & 6.6 & 7.1 & 6.6 & 5.7 & 5.2
\\
 & $\chi^2$/d.o.f. & 38.4/41 & 66.8/55 & 51.8/55 & 25.9/17 & 19.5/22 & 27.9/17 & 13.6/16
\\ 
\hline
 & $\Gamma_{\rm{BMC}}$ & 3.8$^{+0.6}_{-0.7}$ & 3.4$^{+0.8}_{-0.7}$ & 3.8$\pm 0.2$  & 2.6$^{+0.3}_{-0.2}$ & 4.0$^{+1.1}_{-0.5}$ & 2.4$\pm 0.5$ & 4.2$^{+1.4}_{-0.7}$ 
\\
  & $kT_{\rm{s}}$ (eV) & 128$^{+11}_{-19}$ & 131$^{+7}_{-8}$ & 125$\pm 4$ & 110$\pm 11$ & 135$\pm 9$ & 123$^{+9}_{-8}$ & 136$^{+11}_{-10}$
\\
BMC & log$A$ & -0.02$^{+0.02}_{-0.46}$ & -0.54$^{+0.47}_{-0.35}$ & $^{e)}-0.3$ & $^{e)}-0.3$ & $^{e)}-0.3$ & -0.67$^{+0.29}_{-0.25}$ & $^{e)}-0.3$
\\
 & $^{c)}N_{\rm{BMC}}$  & 9.2$\pm 0.5$ & 8.7$\pm 0.3$ & 8.9$^{+0.3}_{-0.4}$ & 8.5$^{+0.7}_{-0.8}$ & 9.0$^{+0.6}_{-0.5}$ & 7.0$\pm 0.6$ & 7.1$\pm 0.6$
\\
 & $^{b)}F_{\rm{BMC}}$  & 7.1 & 6.3 & 6.4 & 7.0 & 6.7 & 5.5 & 5.2
\\
 & $\chi^2$/d.o.f. & 37.5/41 & 68.5/55 & 52.1/56 & 26.7/18 & 17.9/21 & 24.3/17 & 14.0/15
\\
\hline 
\hline
\end{tabular}
\begin{small}
\begin{flushleft}  
{\footnotesize
All fits are obtained with a fixed absorption ($N_H$) of 5.0$\times 10^{20}$ cm$^{-2}$, and all the uncertainties represent a 1$\sigma$ confidence interval for one parameter of interest. Notations from $a)$ -- $e)$ follow the same definitions in Table~\ref{spectrum}.\\
* Different quiescent/outburst stages (in MJD) are marked as the green/orange lines on Fig.~\ref{LC}.\\
%$^{a)}$ Normalisation factor in units of $10^{-5}$\,photons~keV$^{-1}$\,cm$^{-2}$\,s$^{-1}$ at 1\,keV.\\
%$^{b)}$ Unabsorbed flux (0.3--10\,keV) in units of $10^{-13}$\,ergs\,cm$^{-2}$\,s$^{-1}$.\\
%$^{c)}$ Normalisation factor in units of $L_{33}/d^2_{10}$\,ergs\,s$^{-1}$\,kpc$^{-2}$, where $L_{33}$ is in units of $10^{33}$\,ergs\,s$^{-1}$ and $d_{10}$ is in units of 10\,kpc.\\
%$^{d)}$ Uncertainties cannot be reasonably constrained, so logA fixed at -0.3 (see text for details).
%The log$A$ is fixed as -0.3 to denote the 67\% BB emission of the Comptonization spectrum is visible for the Earth observer and other 33\% BB emission is convolved with the Comptonization Green function when its uncertainty cannot be reasonably constrained in the best fit. 
}
\end{flushleft}
\end{small}
\end{table*}

The Hilbert spectrum with the most significant energy passes through the DPS peak as shown in Fig.~\ref{HHT}, and shows a similar pattern to that obtained from the WWZ periodogram.
The most significant Hilbert energy shown on the Hilbert spectrum can be constructed with the HHT of specific components from IMF5 to IMF8 if we decompose the original light curve with an artificial noise level of 0.2 ct/s. 
As shown in the left panel of Fig.~\ref{HHT}, the cycle length of IMF5 is about 90--200\,days, and the typical cycle length of IMF5 corresponds to the duration of each outburst. 
IMF6--8 describe the variation of the recurrence cycle length of each outburst.
Before t$\sim$1400\,d  shown in Fig.~\ref{LC} (2012 Sep.), IMF6 is the most significant signal, which has a modulation of about 1 year. 
IMF7 with a quasi-period of $\sim 450$\,days dominates the instantaneous amplitude from t$\sim$1500\,d--2300\,d (i.e., the end of 2012 to 2015 mid. Feb.), and IMF8 with a cycle length of more than 2 years became more evident from t$\sim$2300\,d (i.e., the end of 2015).

%TTTTTTTTTTTTTTTTTTTTTTTTTTTTTTTTTTTTTTTTTTTTTTTTTTTTTTTTT 
\begin{figure}
\psfig{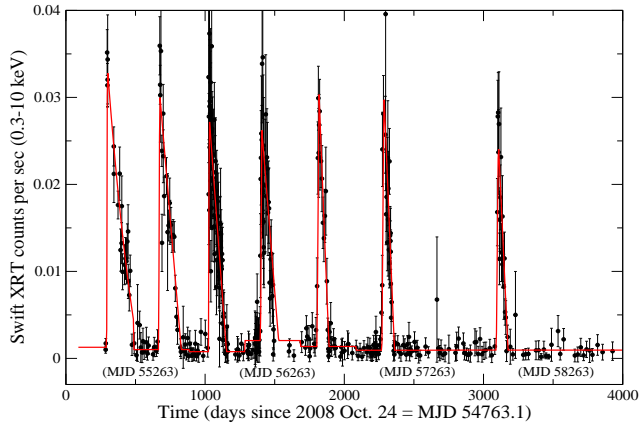}
\caption{\small{Best fit to the \emph{Swift} light curve with a FRED profile. The red FRED profile demonstrates the best fit to the distribution of black data points.}   
}
\label{fitting}
\end{figure}
%%%%%%%%%%%%%%%%%%%%%%%%%%%%%%%%%%%%%%%%%%%%%%%%%%%%  

The non-sinusoidal outburst profile with a rapid rise and an exponential decay can lead to an intra-wave modulation caused by the instantaneous frequency change within one oscillation cycle, and the Hilbert spectrum provides enough resolution to investigate the instability caused by this intra-wave modulation.
For instance, the major signal corresponds to a frequency $\sim$0.002--0.004 d$^{-1}$ which is seen in the Hilbert spectrum over several cycles, and is consistent with the earlier outburst separations.
We note such a signal is highly non-linear, and the variation of instantaneous frequency is dominated by the intra-wave modulation.
We also find that a local signal shows the strongest Hilbert energy with its maximum frequency at an epoch close to the time of peak count rate of each outburst.  
One example is that the signal with a frequency of $\sim$0.0014\,d$^{-1}$ has a peak Hilbert energy at t$\sim$3100\,d (i.e., 2017 Apr. 20), which is comparable to the epoch of the peak luminosity for the 7th outburst.  

The flare candidate with a count rate of 0.007 ct/s detected at t$\sim$2666\,d (i.e., 2016 mid-Feb.; $\sim$MJD 57429) has a separation of $\sim$370 days to the 6th outburst (ref. Fig.~\ref{LC}), and this time interval fits to the separation between the earlier outbursts or the quasi-periodicity determined by the timing analysis via the first five outbursts (from 2009 Aug. to 2013 Dec.; \citealt{Godet2013,Godet2014,Lin2015}).
Nevertheless, we did not detect a continuously increasing flux in the following observations, and the HR determined around the flare is close to 1, which deviates from a soft spectral state usually observed during an outburst.
We also did not find any clear outburst indication related to this candidate event in the dynamical timing analysis, and the large uncertainty (0.007 ct/s) due to the very low count rate makes this flare a statistical fluctuation.
The next (7th) outburst shown in Fig.~\ref{LC} \citep{YY2017} has a separation of $> 800$ days to the previous burst \citep{KSF2015}, which is twice as long as in the earlier outburst intervals.
We used a FRED profile to fit the light curve as shown in Fig.~\ref{fitting}.
Except for the first outburst, the exponential index in the FRED profile determined from the fit is close to 0, and the structure of the profile is similar to a triangular wave.  
Based on the fit, we then determined the epoch of the beginning and end of each outburst, the epoch of each outburst to reach the most luminous state and the highest count rate achieved in each outburst.
HLX$-$1 has experienced 7 major outbursts since 2008, with recurrence intervals of 375.5, 355.5, 375.5, 411.4, 467.3 and 822.8\,d, as determined from the peak fits to a template profile.
We also obtain similar values if we assume a more complicated FRED profile to fit the light curve.
From the aforementioned fit, we determine the length of the duty cycle for the 7 outbursts as $\sim$200, 168, 132, 129, 88, 96 and 84 days.
Except for the 6th outburst, we find that the outburst duration and the fluence continuously decreased.

\subsection{Results obtained from HID and spectral fits}
\label{sec:spectral result}

In Fig.~\ref{HR}, we show that outbursts of HLX$-$1 have a relatively low HR, which corresponds to a soft spectrum shown in the HID.
Most of the HID data points concentrate on both low/hard and high/soft sides, which is similar to the results inferred from earlier studies \citep{Yan2015,TS2016II}.
The HRs usually increased when HLX$-$1 was in the low luminosity state.
Table~\ref{spectrum} clearly demonstrates that the spectra of high-HR bands provide better fits to a single PL component.
If we consider composite models to fit the spectra obtained with HR $>$ 0.5, then the best fit also indicates that the non-thermal flux is about 5 times larger than the thermal contribution.
However, the contribution of the PL component cannot improve the fit to the spectra of low-HR (i.e., `F' and `G') bands.
The spectra determined at low-HR bands have larger photon indices, except for `G' band spectra with BMC model fits, but the associated errors on the PL
are very large.
The spectral parameters obtained from different HR bands can represent how the spectral behavior changed between the low/hard state, steep PL (intermediate) state to high/soft state \citep{Soria2007,Hu2018}.
In the fit to the PL+discBB model, we obtained consistent spectral parameters as shown in \citet{Godet2012} and \citet{Yan2015}.
Compared with the PL+BB model, the PL+discBB model can provide a higher blackbody temperature ($kT_{\rm{in}}$) for spectra in the same HR band.
The mass of HLX$-$1 is comparable to an IMBH of $10^4$\,$M_{\odot}$ if we assume the inner disc size inferred from the spectral analysis as the innermost stable circular orbit (i.e., $R_{\rm{in}}=R_{\rm{ISCO}}\equiv 6\alpha GM/c^2$, where $\alpha$ corresponds to the spin of a BH; $\alpha$ is 1/6 for an extreme Kerr BH and 1 for a Schwarzschild BH).
In the fit with the BMC model to the spectra generated from different HR bands, we obtain a relatively higher seed photon temperature ($kT_{\rm{s}}$) from 110\,eV to 130\,eV, a larger change of photon index ($\Gamma_{\rm{BMC}}$) from 1.2 to 4.4, and a wider range for the BMC normalization (i.e., (0.6 -- 11)$\times L_{33}/d^2_{10}$\,ergs\,s$^{-1}$\,kpc$^{-2}$) in contrast to results in \citet{TS2016II}.
We also individually checked whether the data obtained after the 6th outburst ($\sim$MJD 57120) can lead to this difference; however, the source is too faint to obtain a statistically significant result in spectral analysis.

According to the profile fit to the light curve (cf. Fig.~\ref{fitting}), we can precisely determine the time interval of each outburst and of the quiescent stage between outbursts.
We then accumulate those observations in the same outburst/quiescent interval in order to search for changes in the spectral parameters, as summarised in Table~\ref{ev_spectra}.
In the outburst stage, we only consider those observations with a count rate larger than the average value as labelled in Fig.~\ref{LC} to investigate the spectral behavior.
During each quiescent stage, we used Cash statistics in the spectral fit because the accumulated source photons are few.
Our results in Table~\ref{ev_spectra} confirm a spectral transition between outburst and quiescence.
Furthermore, photon indices in the 2nd and 6th quiescent intervals are harder compared with other quiescent stages and the associated HRs in these two intervals also have larger values (i.e., 0.77 and 0.65).

Most of the spectra obtained in the outburst stage cannot be described with a single component model, and an acceptable fit requires a composite model of PL+BB/discBB or BMC \citep{TMK97}.
The thermal component dominates almost all the spectra obtained in the outburst stage except for the 4th outburst.
In the 4th outburst stage, a single PL component is enough to provide an acceptable fit and the pure blackbody emission cannot correctly describe the spectral behavior.
In addition, we also note that this spectral fit is relatively poor, and the probability given by an F-test to have an additional thermal component to improve the fit (as shown for PL+discBB in Table~\ref{ev_spectra}) is only 0.38.     
For outburst stage spectra described with a PL+BB/discBB model, the outbursts do not demonstrate a significantly softer photon index than that in the following quiescent stage. 
By contrast, the BMC model can describe all the spectra with a much softer photon index in outburst than in quiescence.
Except for the 4th and 6th outbursts, the BB emission of the composite PL+BB model has a blackbody temperature ($kT_{\rm{BB}}$) of about 150 eV, which is higher than that given by the fit to the spectra of low HR bands (i.e., Band D to G) in \citet{TS2016II}.
The BB emission in the BMC model can be described with a temperature ($kT_{\rm{BMC}}$) of $\sim 120-140$\,eV, which is consistent with that determined by the fit to the spectra at a high-soft state \citep{TS2016II}.
We also note the parameter of the Comptonized fraction (log$A$) is difficult to constrain in some outburst spectra, and the spectral fit can be marginally improved with a harder photon index if we fixed log$A$ close to 0. 
The decreasing trend of source flux in recent outbursts can be roughly inferred from our spectral analysis (cf. 4--7th outbursts in Table~\ref{ev_spectra}).

%TTTTTTTTTTTTTTTTTTTTTTTTTTTTTTTTTTTTTTTTTTTTTTTTTTTTTTTTT 
\begin{figure}
\psfig{file=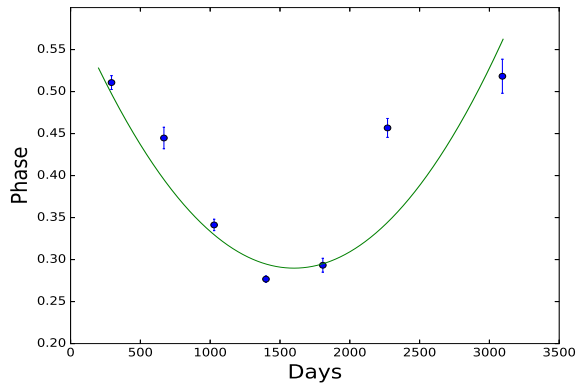,width=3.3in, height=2.2in}
\caption{\small{Phase residual fitting to the HLX$-$1 outburst times, assuming that these events are driven by an orbital modulation. The blue dots indicate the phase of peak epochs determined via a preliminary timing solution with low-order terms, and the green curve shows the fit to the phase residual with an improved solution. The x-axis origin corresponds to MJD~54763.1, as in all other figures.}  
}
\label{phase}
\end{figure}
%%%%%%%%%%%%%%%%%%%%%%%%%%%%%%%%%%%%%%%%%%%%%%%%%%%%  
 
\section{Discussion}
\label{sec:discussion}

We have performed a thorough temporal and spectral analysis of the \emph{Swift} dataset on HLX$-$1, and find that the dominant long-term periodicity is $396\pm 2$\,d.
According to the WWZ periodogram and the Hilbert spectrum shown in Figures~\ref{periodogram} and \ref{HHT}, we see that the main signal gradually decreased in significance after $\sim$MJD 57050 and a new signal with frequency lower than 0.002\,d$^{-1}$ became stronger.

We can derive an ephemeris assuming that the major signal is an orbital period by phase residual fitting to the peak epoch of each outburst.
We use the maximum-likelihood method \citep{Livingstone2009} to determine the phase shift with a fundamental solution of only a fixed orbital period.
As shown in Fig.~\ref{phase}, the solution with only a period cannot align the phase of each peak so we tried to increase a high-order polynomial term in the solution to account for the variation of the orbital period to refit the phase residuals.
Table~\ref{orbital_solution} describes the possible orbital ephemeris with an abnormal period derivative.
Although the fit is still poor, we note that the number of peaks in this fit are too few to provide a more complicated orbital ephemeris (e.g., elliptical orbit), and the solution cannot be significantly improved even with an increase of the second derivative of the orbital period.

The signal caused by an orbital modulation can be due to either partial eclipses by the companion star or enhanced accretion during the periastron passage in an eccentric orbit as in Be/X-ray binaries.
We note that the FRED-like cycle structure detected in the light curve is not similar to the orbital profiles of eclipsing X-ray binaries (see e.g. \citealt{HMSK90,Ioannou2002,Falanga2015,RMP2018}). 
Furthermore, although the quiescent-outburst cycles \citep{Reig2007} and the transition of the spectral behavior \citep{YY2009} are similar to that of Be/X-ray binaries, HLX$-$1 is thought to be an IMBH and a huge orbital period derivative shown in Table~\ref{orbital_solution} clearly demonstrates an instability of this system that contradicts such an explanation for typical Be/X-ray binaries. 	

\citet{Lasota2011} suggest that the observed outburst recurrence can be due to enhanced mass transfer from an AGB star at periastron passage of the IMBH.
The small delay in the rise of an outburst detected between the optical and the X-ray bands potentially indicates a small mass delivery radius in a highly eccentric orbit, and the stochastic fluctuation of the signal can be simulated by a modulated mass transfer due to tidal stripping of a star in this eccentric orbit via smooth particle hydrodynamical simulations. 
\citet{Godet2014} considered different donor structures, equations of state and periapsis separations in a polytrope model to simulate the orbital period evolution due to the effect of mass loss to periapsis passages.
When the orbital parameters have sudden changes from orbit to orbit, the quasi-periodic signal obtained from the simulation seems to satisfy the observed outburst recurrence, including the latest delay of an outburst. 

If we normalize all the outburst separations for HLX$-$1 to 400 days, we obtain a dimensionless time sequence of $\sim$[0.94, 0.89, 0.94, 1.03, 1.17, 2.06] to compare with the simulated result provided in \citet{Godet2014}.
The final cycle in the time sequence has a timescale of more than double that of the earlier cycles, and it potentially indicates that the mass transfer will lead the orbit to increase its eccentricity to ultimately become unbound  \citep{MGRO2013}.
We have not detected a new outburst for more than 840 days and this is already longer than the final cycle of the time sequence.
Though we cannot confirm whether the orbit has become parabolic, the significant delay of an outburst can accompany the process for an orbital eccentricity approaching 1 based on the mass transferring scenario.
In the simulation of the orbital parameters due to mass transfer during multiple periapsis passages \citep{Godet2014}, we note that there can be an increase or decrease of the orbital period of each cycle due to the development of stochastic fluctuations inside the donor by adding or removing orbital energy from the system.
The period fluctuations can be more obvious when an orbit almost becomes unbound, but our observation of HLX$-$1 outburst separations so far reveals only an increasing trend without significant fluctuations.
This can be seen through the major signal in the WWZ map (right panel of Fig~\ref{periodogram}) that shows a gradual decay in the frequency domain without significant fluctuations, and also the variation of the high-resolution Hilbert spectrum (Fig.~\ref{HHT}) is not dominated by the change of cycle length (i.e., inter-wave modulation).
We speculate that more significant and continuous delays for the two new outbursts that occurred after t$\sim$2240\,d (i.e., $\sim$MJD~57000) deviated from the simulation results and therefore challenge the orbital interpretation of the origin of the major timing signal.  

In the profile fit to the X-ray outbursts, we find that the duration of an outburst has continuously decreased to half the original length over the period from 2009 to 2013.
The shorter outburst duration may indicate a shorter time for substantial mass transfer in an orbit with a smaller curvature or a larger eccentricity.     
Following the \citet{Godet2014} model, we infer that the orbit is approaching a parabolic form, as we are seeing a continuous delay in the occurrence of outbursts and their duration is becoming shorter.  
Nevertheless, the two recent outbursts that occurred in 2015 and 2017 both have an outburst duration similar to that of the 2013 outburst (85--95 days), but the separation of the recent two outburst intervals have been significantly extended.  
More investigations of mass transfer behaviour at periapsis from a donor star in an eccentric orbit are required if we would like to interpret the outburst recurrence time as the orbital period.

The spectral state transitions between outburst and quiescence shown in Table~\ref{ev_spectra} and in previous studies \citep{Godet2009,Servillat2011,TS2016II} can be caused by the supercritical accretion occurring during periastron passage, and such an interpretation suggests that the outbursts are orbital in nature.
However, we note that an outburst interval related to the spectral state transitions for X-ray binaries does not represent an orbital period \citep{Belloni2005,ST2009,TS2016I}.
In our spectral analysis, we find that the HRs and the photon indices can also change in the quiescent stage.
This result also weakens the correlation between a sudden change of spectral parameters and orbital effects, so the other crucial evidence such as an additional spectral component related to the supercritical accretion episode is required to strengthen the relation between the outburst interval and the orbital period.
We expect that most of the outflows can switch off and the disc X-ray emission can be directly observed once the accretion rate reduces.
Consequently, a rapid increase of the X-ray and optical luminosity in a super-Eddington accretion episode can serve as a precursor to the main outburst \citep{Godet2014}.
However, the current \emph{Swift} observational strategy cannot allow us to confirm such an event preceding the start of the outburst.

%TTTTTTTTTTTTTTTTTTTTTTTTTTTTTTTTTTTTTTT
\begin{table}
\begin{center}
\caption[]{Orbital ephemeris of \eso\ derived with phase residual fitting (Fig.~\ref{phase}) to \emph{Swift} data.}\label{orbital_solution}
\begin{tabular}{ll}
\hline\hline
\multicolumn{2}{l}{Parameter}\\
\hline
Target name\dotfill & \eso\ \\
Valid MJD range\dotfill & 54853--58043 \\
Right ascension, $\alpha$\dotfill & 1:10:28.25* \\
Declination, $\delta$\dotfill &  $-$46:4:20.3* \\
Orbital period, $P_{orb}$ (days)\dotfill & 354(9) \\
First derivative of orbital period, $\dot{P}_{orb}$ (s\,s$^{-1}$)\dotfill & $3.0(7)\times10^{-2}$ \\
Epoch of frequency determination (MJD)\dotfill & 55048.55423 \\
Time system \dotfill & TDB \\
$\chi_{\nu}^2$/dof\dotfill & 32/4 \\
\hline
\end{tabular}
\begin{small}
\begin{flushleft}
{\footnotesize
*90\% uncertainty is $3.5$\arcsec\ .\\
Note: the numbers in parentheses denote errors in the last digit. The errors of $P_{orb}$ and $\dot{P}_{orb}$ are obtained after multiplying the individual data point errors by a factor that brought the $\chi_{\nu}^2$/dof to one.\\
}
\end{flushleft}
\end{small}
\end{center}
\end{table} 
%TTTTTTTTTTTTTTTTTTTTTTTTTTTTTTTTTTTTTTT 
    
In comparison to the chaotic orbital modulation model, we suggest that the quasi-periodic variation detected in HLX$-$1 is more likely related to a superorbital period.
In this case, the real orbital period of this system should be much shorter \citep{KC2012}. 
Superorbital modulations can occur on a wide range of timescales from tens to hundreds of days, and may originate from occultation by a warped, irradiated accretion disc \citep{OD2001} or tidal-induced precession of a tilted accretion disc  \citep{WK91}.
Because we can detect a significant delay for recent outbursts, the origin of the superorbital modulation is unlikely to be close to a stable switching time-scale between a warped disc and a flat disc (e.g., eq. (2) in \citet{SB2004}).
In addition, the spectral transition/variations are found to be an absorption effect in the warped accretion disc model \citep{Brightman2016}, but such an effect is not seen in our data.
Compared to the warped disc model, precession driven by tidal interaction can only occur when the mass ratio between the companion and the compact star ($q = M_C/M_X$) is smaller than $0.25-0.33$ \citep{WK91} , and an IMBH can easily satisfy this criterion.
The spectral variations attributed to a precessing accretion disc can also be detected \citep{Furst2017} when the angle between the axis of the disc and the line of sight changes.   
We would expect to detect a superorbital modulation in the optical band if there is also significant optical emission from the outer ring of the accretion disc \citep{Hu2019}.     

The outer disc size of a few $10^{13}$\,cm, inferred from optical observations \citep{Soria2012}, provides an alternative explanation to the evolution of the light curve if outbursts of HLX$-$1 were caused by the mass transfer instability arising from the oscillating inner disc wind scenario \citep{Soria2017}.
We note that the amount of mass involved in each outburst duty cycle only contains a small fraction of the mass stored in the disc and the outburst will terminate when the innermost part of the disc is depleted by wind.
When the disc wind is strong enough to regulate the outbursts of HLX$-$1, a damped disc instability occurs \citep{SMLB86} if accretion in the X-ray emitting region triggers a massive wind in an annulus of the disc. 
A positive perturbation of the accretion would push the system over the threshold for the high/soft state, and the subsequent wind-driven oscillation of the accretion would being the system back to a hard state \citep{Soria2017}.
The wind oscillation period can be only a few years \citep{SMLB86} to fit the recurrence time of an outburst for HLX$-$1, and both the delays of outbursts and a reduced fluence in each outburst interval can be explained with damping of the wind oscillations when the wind multiplication factor decreases.
The narrow H$_{\alpha}$ emission line detected from optical spectra \citep{SHP2013} suggests the presence of diffuse gas around our target and provides a marginal evidence for a disc wind.
The wind oscillation mechanism also suggests that the binary period (i.e. 10--12 d) is much shorter than the recurrence period of an outburst in our case; however, the current X-ray observational cadence is longer than the inferred binary period and makes it difficult for us to confirm it from the Fourier-based timing methods.
No new outbursts have been detected since 2017 June (i.e., MJD~57920), which can be a consequence of the damping of the wind oscillation, or the next outburst has significantly been delayed again.  

In order to distinguish between different scenarios for the recurrent outbursts of HLX$-$1, it will be necessary to have X-ray monitoring observations with a higher cadence (i.e., $\leq 2$\,days) and exposures of a few thousand seconds.
This will be essential for detecting the change of luminosity preceding an outburst, as long-term optical monitoring is challenging for such a faint ($\sim$24th magnitude) counterpart \citep{Soria2010}.

\section*{Acknowledgments}

We thank the referee Prof. Phil Charles for his very detailed comments and useful suggestions that helped us improve the manuscript.
This work made use of data supplied by the archival data server of NASA’s High Energy Astrophysics Science Archive Research Center (HEASARC) and  
by the UK \emph{Swift} Science Data Centre at the University of Leicester.
The Korean team in this work is supported by the National Research Foundation of Korea (NRFK) through grant 2016R1A5A1013277. 
Y.M.K. is also supported by the NRFK through grant 2019R1C1C1010571.
C.-P.H. acknowledges support form the Japan Society for the Promotion of Science (JSPS, ID: P18318)
J.T. is supported by National Science Foundation of China (NSFC) grants through grants 11573010, U1631103, U1838102, and 11661161010. 
A.K.H.K. is supported by the Ministry of Science and Technology (MoST) of Taiwan through grants 105-2112-M-007-033-MY2, 105-2119-M-007-028-MY3, and 106-2918-I-007-005. 
D.C.-C.Y. is supported by the MoST of Taiwan through grants 107-2115-M-030-005-MY2.

%\bibliography{ESO243}

\begin{thebibliography}{80}

\bibitem[\protect\citeauthoryear{Abbott, Abbott, Abbott, Abernathy, Acernese,
  Ackley, Adams, Adams, Addesso, Adhikari, Adya, Affeldt, Agathos, Agatsuma,
  Aggarwal \& et. al.}{Abbott et~al.}{2016}]{GW2016}
Abbott B.~P.,  Abbott R.,  Abbott T.~D.,  Abernathy M.~R.,  Acernese F.,
  Ackley K.,  Adams C.,  Adams T.,  Addesso P.,  Adhikari R.~X.,  Adya V.~B.,
  Affeldt C.,  Agathos M.,  Agatsuma K.,  Aggarwal N.,    et. al. 2016, Phys.
  Rev. Lett., 116, 061102

\bibitem[\protect\citeauthoryear{{An}, {Lu} \& {Wang}}{{An}
  et~al.}{2016}]{TLW2016}
{An} T.,  {Lu} X.-L.,    {Wang} J.-Y.,  2016, A\&A, 585, A89

\bibitem[\protect\citeauthoryear{{Bachetti}, {Harrison}, {Walton},
  {Grefenstette}, {Chakrabarty}, {F{\"u}rst}, {Barret}, {Beloborodov}, {Boggs},
  {Christensen}, {Craig}, {Fabian}, {Hailey}, {Hornschemeier} \& {et
  al.}}{{Bachetti} et~al.}{2014}]{Bachetti2014}
{Bachetti} M.,  {Harrison} F.~A.,  {Walton} D.~J.,  {Grefenstette} B.~W.,
  {Chakrabarty} D.,  {F{\"u}rst} F.,  {Barret} D.,  {Beloborodov} A.,  {Boggs}
  S.~E.,  {Christensen} F.~E.,  {Craig} W.~W.,  {Fabian} A.~C.,  {Hailey}
  C.~J.,  {Hornschemeier} A.,    {et al.} 2014, Nature, 514, 202

\bibitem[\protect\citeauthoryear{{Belloni}, {Homan}, {Casella}, {van der Klis},
  {Nespoli}, {Lewin}, {Miller} \& {M{\'e}ndez}}{{Belloni}
  et~al.}{2005}]{Belloni2005}
{Belloni} T.,  {Homan} J.,  {Casella} P.,  {van der Klis} M.,  {Nespoli} E.,
  {Lewin} W.~H.~G.,  {Miller} J.~M.,    {M{\'e}ndez} M.,  2005, A\&A, 440, 207

\bibitem[\protect\citeauthoryear{{Brightman}, {Harrison}, {Walton}, {Fuerst},
  {Hornschemeier}, {Zezas}, {Bachetti}, {Grefenstette}, {Ptak}, {Tendulkar} \&
  {Yukita}}{{Brightman} et~al.}{2016}]{Brightman2016}
{Brightman} M.,  {Harrison} F.,  {Walton} D.~J.,  {Fuerst} F.,  {Hornschemeier}
  A.,  {Zezas} A.,  {Bachetti} M.,  {Grefenstette} B.,  {Ptak} A.,  {Tendulkar}
  S.,    {Yukita} M.,  2016, ApJ, 816, 60

\bibitem[\protect\citeauthoryear{{Brown}}{{Brown}}{1991}]{Brown91}
{Brown} J.~C.,  1991, J. Acoust. Soc. Am., 89, 425

\bibitem[\protect\citeauthoryear{{Cash}}{{Cash}}{1979}]{Cash79}
{Cash} W.,  1979, ApJ, 228, 939

\bibitem[\protect\citeauthoryear{{Chen}, {Shrader} \& {Livio}}{{Chen}
  et~al.}{1997}]{CSL97}
{Chen} W.,  {Shrader} C.~R.,    {Livio} M.,  1997, ApJ, 491, 312

\bibitem[\protect\citeauthoryear{{Clarkson}, {Charles}, {Coe} \&
  {Laycock}}{{Clarkson} et~al.}{2003{\natexlab{b}}}]{Clarkson2003b}
{Clarkson} W.~I.,  {Charles} P.~A.,  {Coe} M.~J.,    {Laycock} S.,  2003,
  MNRAS, 343, 1213

\bibitem[\protect\citeauthoryear{{Clarkson}, {Charles}, {Coe}, {Laycock},
  {Tout} \& {Wilson}}{{Clarkson} et~al.}{2003{\natexlab{a}}}]{Clarkson2003a}
{Clarkson} W.~I.,  {Charles} P.~A.,  {Coe} M.~J.,  {Laycock} S.,  {Tout} M.~D.,
     {Wilson} C.~A.,  2003, MNRAS, 339, 447

\bibitem[\protect\citeauthoryear{{Dunn}, {Fender}, {K{\"o}rding}, {Belloni} \&
  {Cabanac}}{{Dunn} et~al.}{2010}]{Dunn2010}
{Dunn} R.~J.~H.,  {Fender} R.~P.,  {K{\"o}rding} E.~G.,  {Belloni} T.,
  {Cabanac} C.,  2010, MNRAS, 403, 61

\bibitem[\protect\citeauthoryear{{Evans}, {Beardmore}, {Page}, {Osborne},
  {O'Brien}, {Willingale}, {Starling}, {Burrows}, {Godet}, {Vetere}, {Racusin},
  {Goad}, {Wiersema}, {Angelini} \& {et al.}}{{Evans} et~al.}{2009}]{Evans2009}
{Evans} P.~A.,  {Beardmore} A.~P.,  {Page} K.~L.,  {Osborne} J.~P.,  {O'Brien}
  P.~T.,  {Willingale} R.,  {Starling} R.~L.~C.,  {Burrows} D.~N.,  {Godet} O.,
   {Vetere} L.,  {Racusin} J.,  {Goad} M.~R.,  {Wiersema} K.,  {Angelini} L.,
   {et al.} 2009, MNRAS, 397, 1177

\bibitem[\protect\citeauthoryear{{Evans}, {Beardmore}, {Page}, {Tyler},
  {Osborne}, {Goad}, {O'Brien}, {Vetere}, {Racusin}, {Morris}, {Burrows},
  {Capalbi}, {Perri}, {Gehrels} \& {Romano}}{{Evans} et~al.}{2007}]{Evans2007}
{Evans} P.~A.,  {Beardmore} A.~P.,  {Page} K.~L.,  {Tyler} L.~G.,  {Osborne}
  J.~P.,  {Goad} M.~R.,  {O'Brien} P.~T.,  {Vetere} L.,  {Racusin} J.,
  {Morris} D.,  {Burrows} D.~N.,  {Capalbi} M.,  {Perri} M.,  {Gehrels} N.,
  {Romano} P.,  2007, A\&A, 469, 379

\bibitem[\protect\citeauthoryear{{Falanga}, {Bozzo}, {Lutovinov},
  {Bonnet-Bidaud}, {Fetisova} \& {Puls}}{{Falanga} et~al.}{2015}]{Falanga2015}
{Falanga} M.,  {Bozzo} E.,  {Lutovinov} A.,  {Bonnet-Bidaud} J.~M.,  {Fetisova}
  Y.,    {Puls} J.,  2015, A\&A, 577, A130

\bibitem[\protect\citeauthoryear{{Farrell}, {Servillat}, {Pforr}, {Maccarone},
  {Knigge}, {Godet}, {Maraston}, {Webb}, {Barret}, {Gosling}, {Belmont} \&
  {Wiersema}}{{Farrell} et~al.}{2012}]{Farrell2012}
{Farrell} S.~A.,  {Servillat} M.,  {Pforr} J.,  {Maccarone} T.~J.,  {Knigge}
  C.,  {Godet} O.,  {Maraston} C.,  {Webb} N.~A.,  {Barret} D.,  {Gosling}
  A.~J.,  {Belmont} R.,    {Wiersema} K.,  2012, ApJL, 747, L13

\bibitem[\protect\citeauthoryear{{Farrell}, {Webb}, {Barret}, {Godet} \&
  {Rodrigues}}{{Farrell} et~al.}{2009}]{Farrell2009}
{Farrell} S.~A.,  {Webb} N.~A.,  {Barret} D.,  {Godet} O.,    {Rodrigues}
  J.~M.,  2009, Nature, 460, 73

\bibitem[\protect\citeauthoryear{{Foster}}{{Foster}}{1996}]{Foster96}
{Foster} G.,  1996, AJ, 112, 1709

\bibitem[\protect\citeauthoryear{{F{\"u}rst}, {Walton}, {Stern}, {Bachetti},
  {Barret}, {Brightman}, {Harrison} \& {Rana}}{{F{\"u}rst}
  et~al.}{2017}]{Furst2017}
{F{\"u}rst} F.,  {Walton} D.~J.,  {Stern} D.,  {Bachetti} M.,  {Barret} D.,
  {Brightman} M.,  {Harrison} F.~A.,    {Rana} V.,  2017, ApJ, 834, 77

\bibitem[\protect\citeauthoryear{{Godet}, {Barret}, {Webb}, {Farrell} \&
  {Gehrels}}{{Godet} et~al.}{2009}]{Godet2009}
{Godet} O.,  {Barret} D.,  {Webb} N.~A.,  {Farrell} S.~A.,    {Gehrels} N.,
  2009, ApJL, 705, L109

\bibitem[\protect\citeauthoryear{{Godet}, {Lombardi}, {Antonini}, {Barret},
  {Webb}, {Vingless} \& {Thomas}}{{Godet} et~al.}{2014}]{Godet2014}
{Godet} O.,  {Lombardi} J.~C.,  {Antonini} F.,  {Barret} D.,  {Webb} N.~A.,
  {Vingless} J.,    {Thomas} M.,  2014, ApJ, 793, 105

\bibitem[\protect\citeauthoryear{{Godet}, {Plazolles}, {Kawaguchi}, {Lasota},
  {Barret}, {Farrell}, {Braito}, {Servillat}, {Webb} \& {Gehrels}}{{Godet}
  et~al.}{2012}]{Godet2012}
{Godet} O.,  {Plazolles} B.,  {Kawaguchi} T.,  {Lasota} J.-P.,  {Barret} D.,
  {Farrell} S.~A.,  {Braito} V.,  {Servillat} M.,  {Webb} N.,    {Gehrels} N.,
  2012, ApJ, 752, 34

\bibitem[\protect\citeauthoryear{{Godet}, {Webb}, {Barret}, {Farrell},
  {Gehrels}, {Servillat} \& {Soria}}{{Godet} et~al.}{2013}]{Godet2013}
{Godet} O.,  {Webb} N.,  {Barret} D.,  {Farrell} S.,  {Gehrels} N.,
  {Servillat} M.,    {Soria} R.,  2013, The Astronomer's Telegram, 5439

\bibitem[\protect\citeauthoryear{{Han}, {An}, {Wang}, {Lin}, {Xie}, {Xu},
  {Hong} \& {Frey}}{{Han} et~al.}{2012}]{Han2012}
{Han} X.,  {An} T.,  {Wang} J.-Y.,  {Lin} J.-M.,  {Xie} M.-J.,  {Xu} H.-G.,
  {Hong} X.-Y.,    {Frey} S.,  2012, Research in Astronomy and Astrophysics,
  12, 1597

\bibitem[\protect\citeauthoryear{{Hellier}, {Mason}, {Smale} \&
  {Kilkenny}}{{Hellier} et~al.}{1990}]{HMSK90}
{Hellier} C.,  {Mason} K.~O.,  {Smale} A.~P.,    {Kilkenny} D.,  1990, MNRAS,
  244, 39

\bibitem[\protect\citeauthoryear{{Horne} \& {Baliunas}}{{Horne} \&
  {Baliunas}}{1986}]{HB86}
{Horne} J.~H.,  {Baliunas} S.~L.,  1986, ApJ, 302, 757

\bibitem[\protect\citeauthoryear{{Hu}, {Chou}, {Wu}, {Yang} \& {Su}}{{Hu}
  et~al.}{2011}]{Hu2011}
{Hu} C.-P.,  {Chou} Y.,  {Wu} M.-C.,  {Yang} T.-C.,    {Su} Y.-H.,  2011, ApJ,
  740, 67

\bibitem[\protect\citeauthoryear{{Hu}, {Kong}, {Ng} \& {Li}}{{Hu}
  et~al.}{2018}]{Hu2018}
{Hu} C.-P.,  {Kong} A. K.~H.,  {Ng} C.~Y.,    {Li} K.~L.,  2018, ApJ, 864, 64

\bibitem[\protect\citeauthoryear{{Hu}, {Mihara}, {Sugizaki}, {Ueda} \&
  {Enoto}}{{Hu} et~al.}{2019}]{Hu2019}
{Hu} C.-P.,  {Mihara} T.,  {Sugizaki} M.,  {Ueda} Y.,    {Enoto} T.,  2019,
  arXiv e-prints, p. arXiv:1910.00200

\bibitem[\protect\citeauthoryear{{Huang} \& {Long}}{{Huang} \&
  {Long}}{2003}]{HL2003}
{Huang} N.~E.,  {Long} S.~R.,  2003, NASA Patent Pending GSC, 14, 673

\bibitem[\protect\citeauthoryear{{Huang}, {Shen}, {Long}, {Wu}, {Shih},
  {Zheng}, {Yen}, {Tung} \& {Liu}}{{Huang} et~al.}{1998}]{Huang98}
{Huang} N.~E.,  {Shen} Z.,  {Long} S.~R.,  {Wu} M.~C.,  {Shih} H.~H.,  {Zheng}
  Q.,  {Yen} N.-C.,  {Tung} C.~C.,    {Liu} H.~H.,  1998, Royal Society of
  London Proceedings Series A, 454, 903

\bibitem[\protect\citeauthoryear{{Ioannou}, {Naylor}, {Smale}, {Charles} \&
  {Mukai}}{{Ioannou} et~al.}{2002}]{Ioannou2002}
{Ioannou} Z.,  {Naylor} T.,  {Smale} A.~P.,  {Charles} P.~A.,    {Mukai} K.,
  2002, A\&A, 382, 130

\bibitem[\protect\citeauthoryear{{Kong}, {Soria} \& {Farrell}}{{Kong}
  et~al.}{2015}]{KSF2015}
{Kong} A.~K.~H.,  {Soria} R.,    {Farrell} S.,  2015, The Astronomer's
  Telegram, 6916

\bibitem[\protect\citeauthoryear{{Kotze} \& {Charles}}{{Kotze} \&
  {Charles}}{2012}]{KC2012}
{Kotze} M.~M.,  {Charles} P.~A.,  2012, MNRAS, 420, 1575

\bibitem[\protect\citeauthoryear{{Lasota}, {Alexander}, {Dubus}, {Barret},
  {Farrell}, {Gehrels}, {Godet} \& {Webb}}{{Lasota} et~al.}{2011}]{Lasota2011}
{Lasota} J.-P.,  {Alexander} T.,  {Dubus} G.,  {Barret} D.,  {Farrell} S.~A.,
  {Gehrels} N.,  {Godet} O.,    {Webb} N.~A.,  2011, ApJ, 735, 89

\bibitem[\protect\citeauthoryear{{Lasota}, {King} \& {Dubus}}{{Lasota}
  et~al.}{2015}]{Lasota2015}
{Lasota} J.-P.,  {King} A.~R.,    {Dubus} G.,  2015, ApJL, 801, L4

\bibitem[\protect\citeauthoryear{{Levine}, {Bradt}, {Chakrabarty}, {Corbet} \&
  {Harris}}{{Levine} et~al.}{2011}]{Levine2011}
{Levine} A.~M.,  {Bradt} H.~V.,  {Chakrabarty} D.,  {Corbet} R.~H.~D.,
  {Harris} R.~J.,  2011, ApJS, 196, 6

\bibitem[\protect\citeauthoryear{{Lin}, {Hu}, {Kong}, {Yen}, {Takata} \&
  {Chou}}{{Lin} et~al.}{2015}]{Lin2015}
{Lin} L.~C.-C.,  {Hu} C.-P.,  {Kong} A.~K.~H.,  {Yen} D.~C.-C.,  {Takata} J.,
   {Chou} Y.,  2015, MNRAS, 454, 1644

\bibitem[\protect\citeauthoryear{{Liu}, {Bregman}, {Bai}, {Justham} \&
  {Crowther}}{{Liu} et~al.}{2013}]{Liu2013}
{Liu} J.-F.,  {Bregman} J.~N.,  {Bai} Y.,  {Justham} S.,    {Crowther} P.,
  2013, Nature, 503, 500

\bibitem[\protect\citeauthoryear{{Livingstone}, {Ransom}, {Camilo}, {Kaspi},
  {Lyne}, {Kramer} \& {Stairs}}{{Livingstone} et~al.}{2009}]{Livingstone2009}
{Livingstone} M.~A.,  {Ransom} S.~M.,  {Camilo} F.,  {Kaspi} V.~M.,  {Lyne}
  A.~G.,  {Kramer} M.,    {Stairs} I.~H.,  2009, ApJ, 706, 1163

\bibitem[\protect\citeauthoryear{{Lomb}}{{Lomb}}{1976}]{Lomb76}
{Lomb} N.~R.,  1976, Ap\&SS, 39, 447

\bibitem[\protect\citeauthoryear{{Makishima}, {Kubota}, {Mizuno}, {Ohnishi},
  {Tashiro}, {Aruga}, {Asai}, {Dotani}, {Mitsuda}, {Ueda}, {Uno}, {Yamaoka},
  {Ebisawa}, {Kohmura} \& {Okada}}{{Makishima} et~al.}{2000}]{Makishima2000}
{Makishima} K.,  {Kubota} A.,  {Mizuno} T.,  {Ohnishi} T.,  {Tashiro} M.,
  {Aruga} Y.,  {Asai} K.,  {Dotani} T.,  {Mitsuda} K.,  {Ueda} Y.,  {Uno} S.,
  {Yamaoka} K.,  {Ebisawa} K.,  {Kohmura} Y.,    {Okada} K.,  2000, ApJ, 535,
  632

\bibitem[\protect\citeauthoryear{{Manukian}, {Guillochon}, {Ramirez-Ruiz} \&
  {O'Leary}}{{Manukian} et~al.}{2013}]{MGRO2013}
{Manukian} H.,  {Guillochon} J.,  {Ramirez-Ruiz} E.,    {O'Leary} R.~M.,  2013,
  ApJL, 771, L28

\bibitem[\protect\citeauthoryear{{Mapelli}, {Annibali}, {Zampieri} \&
  {Soria}}{{Mapelli} et~al.}{2013a}]{MAZS2013I}
{Mapelli} M.,  {Annibali} F.,  {Zampieri} L.,    {Soria} R.,  2013a, A\&A, 559,
  A124

\bibitem[\protect\citeauthoryear{{Mapelli}, {Annibali}, {Zampieri} \&
  {Soria}}{{Mapelli} et~al.}{2013b}]{MAZS2013II}
{Mapelli} M.,  {Annibali} F.,  {Zampieri} L.,    {Soria} R.,  2013b, MNRAS,
  433, 849

\bibitem[\protect\citeauthoryear{{Mapelli}, {Zampieri} \& {Mayer}}{{Mapelli}
  et~al.}{2012}]{MZM2012}
{Mapelli} M.,  {Zampieri} L.,    {Mayer} L.,  2012, MNRAS, 423, 1309

\bibitem[\protect\citeauthoryear{{Miller} \& {Colbert}}{{Miller} \&
  {Colbert}}{2004}]{MC2004}
{Miller} M.~C.,  {Colbert} E.~J.~M.,  2004, International Journal of Modern
  Physics D, 13, 1

\bibitem[\protect\citeauthoryear{{Morrison} \& {McCammon}}{{Morrison} \&
  {McCammon}}{1983}]{MM83}
{Morrison} R.,  {McCammon} D.,  1983, ApJ, 270, 119

\bibitem[\protect\citeauthoryear{{Mu{\~n}oz-Darias}, {Fender}, {Motta} \&
  {Belloni}}{{Mu{\~n}oz-Darias} et~al.}{2014}]{Munoz2014}
{Mu{\~n}oz-Darias} T.,  {Fender} R.~P.,  {Motta} S.~E.,    {Belloni} T.~M.,
  2014, MNRAS, 443, 3270

\bibitem[\protect\citeauthoryear{{Mukai}, {Still}, {Corbet}, {Kuntz} \&
  {Barnard}}{{Mukai} et~al.}{2005}]{Mukai2005}
{Mukai} K.,  {Still} M.,  {Corbet} R.~H.~D.,  {Kuntz} K.~D.,    {Barnard} R.,
  2005, ApJ, 634, 1085

\bibitem[\protect\citeauthoryear{{Musaeva}, {Koribalski}, {Farrell}, {Sadler},
  {Servillat}, {Jurek}, {Lenc}, {Starling}, {Webb}, {Godet}, {Combes} \&
  {Barret}}{{Musaeva} et~al.}{2015}]{Musaeva2015}
{Musaeva} A.,  {Koribalski} B.~S.,  {Farrell} S.~A.,  {Sadler} E.~M.,
  {Servillat} M.,  {Jurek} R.,  {Lenc} E.,  {Starling} R.~L.~C.,  {Webb} N.~A.,
   {Godet} O.,  {Combes} F.,    {Barret} D.,  2015, MNRAS, 447, 1951

\bibitem[\protect\citeauthoryear{{Ogilvie} \& {Dubus}}{{Ogilvie} \&
  {Dubus}}{2001}]{OD2001}
{Ogilvie} G.~I.,  {Dubus} G.,  2001, MNRAS, 320, 485

\bibitem[\protect\citeauthoryear{{Raman}, {Maitra} \& {Paul}}{{Raman}
  et~al.}{2018}]{RMP2018}
{Raman} G.,  {Maitra} C.,    {Paul} B.,  2018, MNRAS, 477, 5358

\bibitem[\protect\citeauthoryear{{Reig}}{{Reig}}{2007}]{Reig2007}
{Reig} P.,  2007, MNRAS, 377, 867

\bibitem[\protect\citeauthoryear{{Scargle}}{{Scargle}}{1982}]{Scargle82}
{Scargle} J.~D.,  1982, ApJ, 263, 835

\bibitem[\protect\citeauthoryear{{Schulz} \& {Mudelsee}}{{Schulz} \&
  {Mudelsee}}{2002}]{SM2002}
{Schulz} M.,  {Mudelsee} M.,  2002, Comput. Geosci., 28, 421

\bibitem[\protect\citeauthoryear{Schulz \& Stattegger}{Schulz \&
  Stattegger}{1997}]{SS1997}
Schulz M.,  Stattegger K.,  1997, Computers \& Geosciences, 23, 929

\bibitem[\protect\citeauthoryear{{Seifina} \& {Titarchuk}}{{Seifina} \&
  {Titarchuk}}{2011}]{ST2011}
{Seifina} E.,  {Titarchuk} L.,  2011, ApJ, 738, 128

\bibitem[\protect\citeauthoryear{{Seifina}, {Titarchuk}, {Shrader} \&
  {Shaposhnikov}}{{Seifina} et~al.}{2015}]{Seifina2015}
{Seifina} E.,  {Titarchuk} L.,  {Shrader} C.,    {Shaposhnikov} N.,  2015, ApJ,
  808, 142

\bibitem[\protect\citeauthoryear{{Servillat}, {Farrell}, {Lin}, {Godet},
  {Barret} \& {Webb}}{{Servillat} et~al.}{2011}]{Servillat2011}
{Servillat} M.,  {Farrell} S.~A.,  {Lin} D.,  {Godet} O.,  {Barret} D.,
  {Webb} N.~A.,  2011, ApJ, 743, 6

\bibitem[\protect\citeauthoryear{{Shaposhnikov} \& {Titarchuk}}{{Shaposhnikov}
  \& {Titarchuk}}{2009}]{ST2009}
{Shaposhnikov} N.,  {Titarchuk} L.,  2009, ApJ, 699, 453

\bibitem[\protect\citeauthoryear{{Shields}, {McKee}, {Lin} \&
  {Begelman}}{{Shields} et~al.}{1986}]{SMLB86}
{Shields} G.~A.,  {McKee} C.~F.,  {Lin} D.~N.~C.,    {Begelman} M.~C.,  1986,
  ApJ, 306, 90

\bibitem[\protect\citeauthoryear{{Soria}}{{Soria}}{2007}]{Soria2007}
{Soria} R.,  2007, Astrophys. Space Sci., 311, 213

\bibitem[\protect\citeauthoryear{{Soria}, {Hakala}, {Hau}, {Gladstone} \&
  {Kong}}{{Soria} et~al.}{2012}]{Soria2012}
{Soria} R.,  {Hakala} P.~J.,  {Hau} G. K.~T.,  {Gladstone} J.~C.,    {Kong} A.
  K.~H.,  2012, MNRAS, 420, 3599

\bibitem[\protect\citeauthoryear{{Soria}, {Hau}, {Graham}, {Kong}, {Kuin},
  {Li}, {Liu} \& {Wu}}{{Soria} et~al.}{2010}]{Soria2010}
{Soria} R.,  {Hau} G. K.~T.,  {Graham} A.~W.,  {Kong} A. K.~H.,  {Kuin} N.
  P.~M.,  {Li} I.~H.,  {Liu} J.-F.,    {Wu} K.,  2010, MNRAS, 405, 870

\bibitem[\protect\citeauthoryear{{Soria}, {Hau} \& {Pakull}}{{Soria}
  et~al.}{2013}]{SHP2013}
{Soria} R.,  {Hau} G.~K.~T.,    {Pakull} M.~W.,  2013, ApJL, 768, L22

\bibitem[\protect\citeauthoryear{{Soria}, {Musaeva}, {Wu}, {Zampieri},
  {Federle}, {Urquhart}, {van der Helm} \& {Farrell}}{{Soria}
  et~al.}{2017}]{Soria2017}
{Soria} R.,  {Musaeva} A.,  {Wu} K.,  {Zampieri} L.,  {Federle} S.,  {Urquhart}
  R.,  {van der Helm} E.,    {Farrell} S.,  2017, MNRAS, 469, 886

\bibitem[\protect\citeauthoryear{{Still} \& {Boyd}}{{Still} \&
  {Boyd}}{2004}]{SB2004}
{Still} M.,  {Boyd} P.,  2004, ApJL, 606, L135

\bibitem[\protect\citeauthoryear{{The LIGO Scientific Collaboration}, {the
  Virgo Collaboration}, {Abbott}, {Abbott}, {Abbott}, {Abraham}, {Acernese},
  {Ackley}, {Adams}, {Adhikari} \& et al.}{{The LIGO Scientific Collaboration}
  et~al.}{2018}]{2018LIGO}
{The LIGO Scientific Collaboration} {the Virgo Collaboration} {Abbott} B.~P.,
  {Abbott} R.,  {Abbott} T.~D.,  {Abraham} S.,  {Acernese} F.,  {Ackley} K.,
  {Adams} C.,  {Adhikari} R.~X.,    et al. 2018, arXiv e-prints

\bibitem[\protect\citeauthoryear{{Titarchuk}, {Mastichiadis} \&
  {Kylafis}}{{Titarchuk} et~al.}{1997}]{TMK97}
{Titarchuk} L.,  {Mastichiadis} A.,    {Kylafis} N.~D.,  1997, ApJ, 487, 834

\bibitem[\protect\citeauthoryear{{Titarchuk} \& {Seifina}}{{Titarchuk} \&
  {Seifina}}{2016a}]{TS2016II}
{Titarchuk} L.,  {Seifina} E.,  2016a, A\&A, 595, A101

\bibitem[\protect\citeauthoryear{{Titarchuk} \& {Seifina}}{{Titarchuk} \&
  {Seifina}}{2016b}]{TS2016I}
{Titarchuk} L.,  {Seifina} E.,  2016b, A\&A, 585, A94

\bibitem[\protect\citeauthoryear{{Webb}, {Cseh}, {Lenc}, {Godet}, {Barret},
  {Corbel}, {Farrell}, {Fender}, {Gehrels} \& {Heywood}}{{Webb}
  et~al.}{2012}]{Webb2012}
{Webb} N.,  {Cseh} D.,  {Lenc} E.,  {Godet} O.,  {Barret} D.,  {Corbel} S.,
  {Farrell} S.,  {Fender} R.,  {Gehrels} N.,    {Heywood} I.,  2012, Science,
  337, 554

\bibitem[\protect\citeauthoryear{{Webb}, {Barret}, {Godet}, {Servillat},
  {Farrell} \& {Oates}}{{Webb} et~al.}{2010}]{Webb2010}
{Webb} N.~A.,  {Barret} D.,  {Godet} O.,  {Servillat} M.,  {Farrell} S.~A.,
  {Oates} S.~R.,  2010, ApJL, 712, L107

\bibitem[\protect\citeauthoryear{{Whitehurst} \& {King}}{{Whitehurst} \&
  {King}}{1991}]{WK91}
{Whitehurst} R.,  {King} A.,  1991, MNRAS, 249, 25

\bibitem[\protect\citeauthoryear{{Wiersema}, {Farrell}, {Webb}, {Servillat},
  {Maccarone}, {Barret} \& {Godet}}{{Wiersema} et~al.}{2010}]{Wiersema2010}
{Wiersema} K.,  {Farrell} S.~A.,  {Webb} N.~A.,  {Servillat} M.,  {Maccarone}
  T.~J.,  {Barret} D.,    {Godet} O.,  2010, ApJL, 721, L102

\bibitem[\protect\citeauthoryear{Wu \& Huang}{Wu \& Huang}{2009}]{Wu2009}
Wu Z.,  Huang N.~E.,  2009, Advances in Adaptive Data Analysis, 1, 1

\bibitem[\protect\citeauthoryear{{Yan} \& {Yu}}{{Yan} \& {Yu}}{2015}]{YY2015}
{Yan} Z.,  {Yu} W.,  2015, ApJ, 805, 87

\bibitem[\protect\citeauthoryear{{Yan} \& {Yu}}{{Yan} \& {Yu}}{2017}]{YY2017}
{Yan} Z.,  {Yu} W.,  2017, The Astronomer's Telegram, 10289

\bibitem[\protect\citeauthoryear{{Yan}, {Zhang}, {Soria}, {Altamirano} \&
  {Yu}}{{Yan} et~al.}{2015}]{Yan2015}
{Yan} Z.,  {Zhang} W.,  {Soria} R.,  {Altamirano} D.,    {Yu} W.,  2015, ApJ,
  811, 23

\bibitem[\protect\citeauthoryear{{Yu} \& {Yan}}{{Yu} \& {Yan}}{2009}]{YY2009}
{Yu} W.,  {Yan} Z.,  2009, ApJ, 701, 1940

\end{thebibliography}
%\bibliographystyle{mn2e}

\label{lastpage}

\end{document}